\documentclass[a4paper, notitlepage]{article}
\usepackage{natbib}
\usepackage{graphicx}
\usepackage[colorlinks=true]{hyperref}
\providecommand{\grll}{Geophys. Res. Lett.}

\providecommand{\affil}[1]{\textsuperscript{#1}}
\providecommand{\affiliation}[2]{{#1}: {#2}\\}
\usepackage{geometry}
\newcommand{\RS}{\ensuremath{\mathrm{R_S}}}
\newcommand{\ssr}{Space Sci. Rev.}
\newcommand{\mnras}{Mon. Not. R. Astro. Soc.}
\usepackage{amsmath}

\pdfoutput=1
\makeatletter
\renewcommand{\maketitle}{
    \begin{center}
      \large
      {\LARGE\@title}
      \par\vspace{1ex}
        \@author
	\par\vspace{1ex}
      \@date
    \end{center}
    \@thanks
}
\makeatother
\begin{document}

\title{The Structure of Planetary Period Oscillations in Saturn's Equatorial Magnetosphere: Results from the Cassini Mission}

\author{\small D. J. Andrews\affil{1},
S. W. H. Cowley\affil{2},
G. Provan\affil{2},
G. J. Hunt\affil{3},
L. Z. Hadid\affil{1},
M. W. Morooka\affil{1},
and J.-E. Wahlund\affil{1}.
\\
\vspace{11pt}
\mbox{Corresponding author: David J. Andrews}, Swedish Institute of Space Physics (Uppsala), Box 537, Uppsala 75121, Sweden.\\
 david.andrews@irfu.se\\
 \vspace{11pt}
 \affiliation{1}{Swedish Institute of Space Physics, Uppsala, Sweden}
 \affiliation{2}{Department of Physics and Astronomy, University of Leicester, UK}
 \affiliation{3}{Blackett Laboratory, Imperial College London, UK.}
}

\date{\vspace{11pt}
September 2019\\ Preprint accepted for publication in \\J. Geophys. Res. (Space Physics)\\}

\onecolumn

\maketitle

\begin{abstract}

Saturn's magnetospheric magnetic field, planetary radio emissions, plasma populations and magnetospheric structure are all known to be modulated at periods close to the assumed rotation period of the planetary interior.
These oscillations are readily apparent despite the high degree of axi-symmetry in the internally produced magnetic field of the planet, and have different rotation periods in the northern and southern hemispheres.
In this paper we study the spatial structure of (near-) planetary period magnetic field oscillations in Saturn's equatorial magnetosphere.
Extending previous analyses of these phenomena, we include all suitable data from the entire Cassini mission during its orbital tour of the planet, so as to be able to quantify both the amplitude and phase of these field oscillations throughout Saturn's equatorial plane, to distances of 30 planetary radii.
We study the structure of these field oscillations in view of both independently rotating northern and southern systems, finding spatial variations in both magnetic fields and inferred currents flowing north-south  that are common to both systems.
With the greatly expanded coverage of the equatorial plane achieved during the latter years of the mission, we are able to present a complete survey of dawn-dusk and day-night asymmetries in the structure of the oscillating fields and currents.
We show that the general structure of the rotating currents is simpler than previously reported, and that the relatively enhanced nightside equatorial fields and currents are due in part to related periodic vertical motion of Saturn's magnetotail current sheet.\vspace{11pt}\\
{\textbf{Plain Language Summary: }}
Saturn's magnetic field, produced in its interior by dynamo processes, is apparently perfectly symmetric with respect to the spin-axis of the planet.
Despite this, measurements of magnetic fields in Saturn's magnetosphere, radio emissions and plasma populations all show oscillations with periods close to the (inferred) rotation rate of the interior of the planet.
The origin of these oscillations is yet to be fully explained, but electrodynamic coupling between the upper atmosphere/ionosphere, and magnetosphere plays a central role.
In this study, we use magnetic field measurements from NASA's Cassini spacecraft to statistically study the spatial structure of the magnetic field oscillations, and the electrical currents producing them.
Expanding on previous studies, we find good overall agreement with existing theoretical models, but with discrepancies suggestive of the influence of solar-wind magnetosphere coupling on the system.\vspace{11pt}\\
{\textbf{Key Points: }}
\begin{itemize}
  \item We conduct a complete analysis of the structure of Saturn's equatorial magnetic field oscillations
  \item Seasonal and beat-phase variations are accounted for in an analysis of dual period oscillations
  \item Electrical currents flow vertically through the equatorial plane in two linked spirals
\end{itemize}
\end{abstract}
\clearpage
\section{Introduction}

Measurements made at Saturn using data from NASA's Cassini orbiter have elucidated a planetary magnetosphere filled with low frequency waves and modulations in plasmas, magnetic fields, and radio and auroral emissions~\citep[see e.g.][for a review of this topic]{carbary13a}.
These rotating phenomena all exhibit rotation periods close to, but typically larger than, the assumed rotation period of the deep interior of the planet of $\sim$10.55~h~\citep{anderson07a, helled15a}, hence the widely used term `planetary period oscillations' (PPO).
The majority of these phenomena further exhibit a specifically $m=1$ rotational modulation at this period, i.e.\ with phase fronts moving through the magnetosphere in the same sense as planetary rotation, at least to first order (where $m$ is the azimuthal wavenumber).
Explaining the origin of these periodic plasma phenomena in Saturn's otherwise apparently perfectly axi-symmetric magnetic field  has been a major focus of the Cassini mission (the mis-alignment between the spin and magnetic axis being less than 0.01$^\circ$~\citep{dougherty18a}).

The presence of these oscillations was first detected in radio data obtained by Voyager~\citep{kaiser80a}, in which the intensity of the Saturnian kilometric radiation (SKR) was found to be modulated at a period of $\sim$10.67~h. Subsequent measurements first with Ulysses and later Cassini have shown both that the period of this modulation drifts on timescales comparable to the shifting Saturn seasons~\citep{kurth07a, kurth08a}, and moreover that SKR emissions from high latitude fields lines over opposing ionospheres are modulated at distinctly different periods~\citep{galopeau00a, gurnett09a, gurnett10b, lamy11a, fischer15a}.
Independent determination of the rotation periods of SKR emissions from the northern and southern hemispheres is possible owing to the dominance of emissions in the R-X radio mode, and the consequent left-hand polarization of emissions from the southern hemisphere, and right-hand from the northern~\citep[e.g.,][]{lamy08a, lamy11a, lamy17a, fischer09a, ye18a}.

Much progress has been made in understanding the causal relations between these different oscillating quantities through development of both empirically-based and theoretically-based models of rotating magnetosphere-ionosphere coupling currents~\citep[e.g.,][]{southwood07a, gurnett07a, provan09a, andrews08a, andrews10b, khurana09a, smith11a, southwood14a}.
In each proposed model, auroral currents link the high-latitude ionosphere of the planet to the equatorial magnetosphere, and may close both across the field through the equatorial plane and via the opposing ionosphere.
Magnetic field perturbations associated with these rotating currents are measurable throughout the magnetosphere, with amplitudes typically in the range of $\sim$0.1 to a few nT~\citep{andrews08a, andrews10a}.
Postulated `sources' of these modulated currents variously lie in systems of atmospheric neutral winds at ionospheric altitudes or stable magnetospheric convection cells~\citep[e.g.,][]{southwood14a, smith16a, andrews10b, jia12b}.

In more detail, seasonal variations in both the period of the two systems and their relative amplitudes have been tracked throughout the Cassini mission (2004-2017), representing the interval from Saturn's southern hemisphere summer through vernal equinox and into northern hemisphere summer.
During the early part of the Cassini mission, prior to equinox, the southern system oscillations were dominant over the northern, and had a clearly separated, longer rotation period~\citep{kurth07a, kurth08a, andrews08a, provan09a}.
Following this interval, the general trend has been a shift away from the dominance of radio emissions and field oscillations associated with the southern system, towards dominance of the corresponding northern system oscillations following equinox~\citep{gurnett10b, gurnett11b, andrews10a, andrews12a, provan11a, provan13a, lamy11a, lamy17a, fischer15a, ye16a, ye18a}.
However, the variations in both relative amplitudes and modulation periods of the two systems have not been steady, instead proceeding via apparent rapid shifts in both amplitude and periodicity, including restricted intervals in which the general trend is halted or even reversed~\citep{provan13a, provan14a, provan16a, provan18a}.

In this paper, we extend the analysis of the spatial structure of the equatorial PPO magnetic field perturbations conducted previously by~\citet{andrews10a}.
In doing so, we significantly increase the volume of data analysed and improve the spatial coverage afforded, particularly in respect to local time (LT) through the afternoon sector.
Furthermore, we update the methodology used to reflect subsequent developments in regards to the simultaneous presence of magnetic field oscillations associated with both the southern and northern PPO systems.
In doing so, we determine the spatial variations of the magnetic field oscillations, under the assumption of a common spatial structure for both the northern and southern systems.
Following this, we are able to fully characterise the system of electrical currents flowing normal to the equatorial plane associated with these oscillations, calculated using a simple numerical approach.

The paper is structured as follows.
We proceed by summarising seasonal variations in the amplitudes and periods of the central `core' region PPO magnetic field perturbations in section~\ref{sec:season}.
The selection of suitable equatorial field data for analysis based on Cassini's orbital tour of Saturn is discussed in section~\ref{sec:proc}.
In section~\ref{sec:field}, salient aspects of the structure of the equatorial PPO fields produced by the simple theoretical model are reviewed, before the full analysis of the equatorial field data themselves is given in section~\ref{sec:eq}.
Inferred north-south current densities are calculated in section~\ref{sec:cur}.
Illustrative examples of the superposed northern and southern PPO oscillations are then presented at three different mission intervals in section~\ref{sec:combined}.
Finally, we discuss implications of some of the features found in the analysis in section~\ref{sec:discussion} and briefly summarize the central results in section~\ref{sec:summary}.

\section{Seasonal variations of PPO fields}\label{sec:season}
\begin{figure}[htp]
  \includegraphics[width=0.8\textwidth]{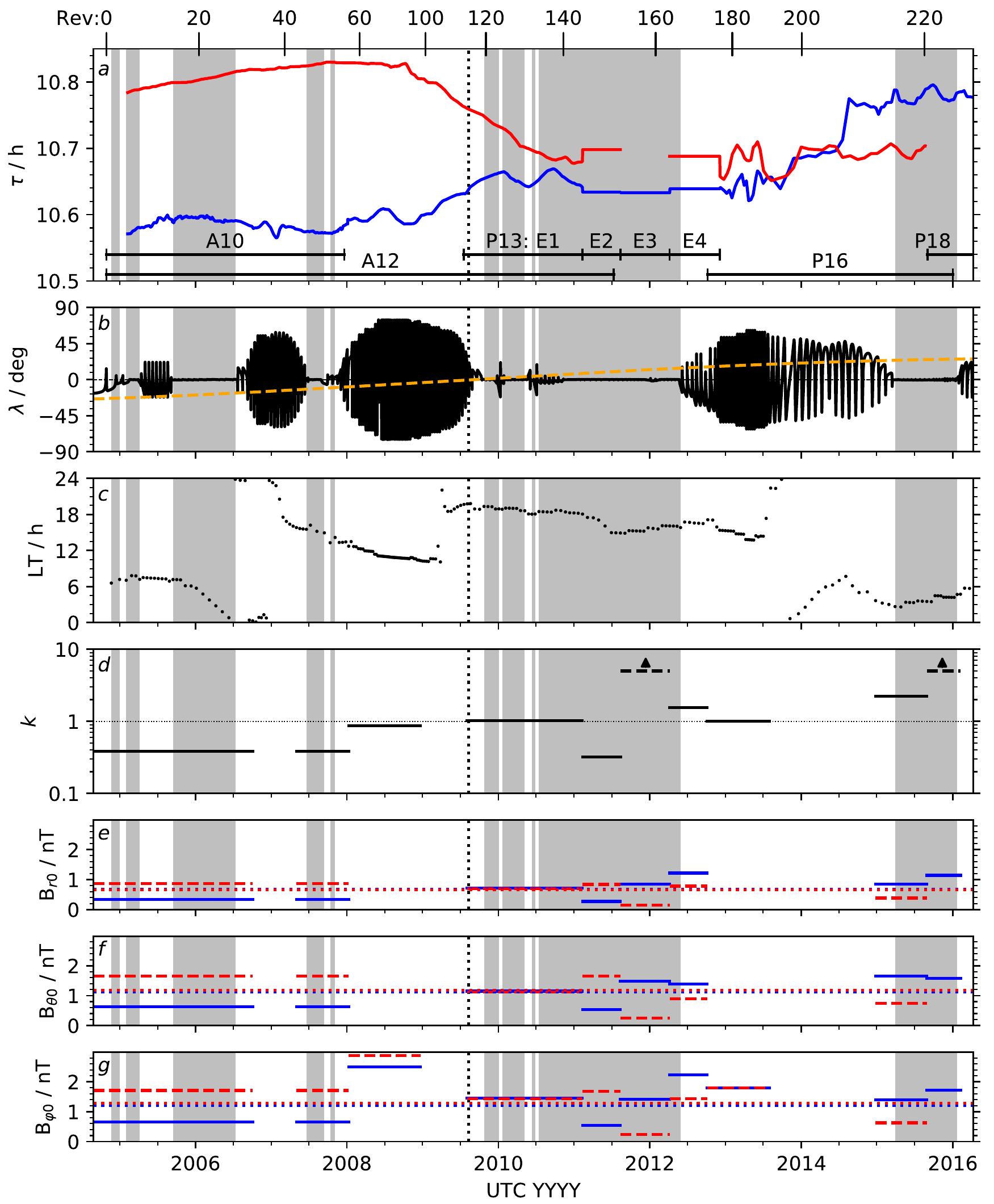}
\caption{PPO properties determined in the series of papers by~\citet{andrews12a}, \citet{provan13a}, \citet{provan16a}, and  \citet{provan18a}.
a) Northern (blue) and southern (red) system rotation periods. The spans of the data analysed in each relevant paper are indicated in this panel by the horizontal labelled bars (with the results of~\citet{provan13a} further sub-divided into intervals E1-E4, and the interval studied by~\citet{andrews10a} also shown.)
b) Planetographic latitude of Cassini (solid black line), and the Sun (orange dashed line).
c) LT of apoapsis for each orbit.
d) Ratio of the northern and southern PPO field amplitudes, $k$, where intervals with lower limit values of $k=5$ shown by the dashed lines and arrows.
e, f, g) Core region PPO amplitudes $B_{iN\!,S0}$ for the $r$, $\theta$ and $\varphi$ components, respectively.  Dashed red lines indicate southern system values, solid blue northern.
Horizontal red and blue dotted lines in each panel indicate the averaged core region amplitudes $\overline{B}_{iN\!,S0}$ described in the text and given in Table~\ref{tab:stuff}.
 Cassini `Rev' numbers are indicated at the top of the figure.
 Those near-equatorial orbits from which magnetic field data are used in this study are indicated by the vertical grey bars within each panel. Vernal equinox is indicated by the vertical dotted line in each panel.}
\label{fig:ts}
 \end{figure}

We begin by summarising the long-term variations in the PPO properties, as determined over the course of the mission in a series of related papers~\citep{andrews08a, andrews12a, provan13a, provan16a, provan18a}.

In Figure~\ref{fig:ts} we show variations in the periods and amplitudes of magnetic field oscillations throughout the interval from Saturn orbit insertion on 1 July 2004, through to the end of the mission on 15 September 2017.
Rotation periods $\tau_N$ and $\tau_S$ obtained from analysis of magnetic field data are shown in Figure~\ref{fig:ts}a.
Briefly, within appropriately sized 100-200 day intervals of data, best-fit rotation periods are obtained such that the variance in oscillation phase residuals is minimised across the interval.
Full details of this approach can be found for example in~\citet{provan16a}.
The southern system rotation period is shown in red, the northern in blue.
Saturn's vernal equinox on 11 August 2009 is marked by the vertical dotted line.
Following this, large, discrete variations in the relative amplitudes of the northern and southern PPO fields were detected and consequently, only piecewise determinations of the northern and southern periods were possible during the interval from 10~February 2011 to 2~December 2012.
Alternative expressions for the time-variable rotation periods of the northern and southern PPO systems are available, most recently the ``Saturn Longitude System-5'', SLS-5, ~\citep{ye18a}, derived from SKR measurements.
However, we note that only minor changes in the results presented in this paper are expected should one of these alternative schemes be used.
Detailed comparisons of the (generally small) differences between these alternative measurements of rotation periods are given by~\cite{provan14a} and \cite{provan16a}.

Figure~\ref{fig:ts}b shows the latitude of Cassini throughout the period of this study, clearly seen to be broken into distinct near-equatorial phases interposed with high-latitude excursions.
The sub-solar latitude is shown by the dashed orange line in this panel, indicating the progression of the seasons throughout the interval plotted.
Cassini's LT of apoapsis is shown in Figure~\ref{fig:ts}c, thus indicating the variation in sampling of LT throughout the mission.

In parallel with determinations of the northern and southern rotation periods,~\citet{andrews12a}, \citet{provan13a}, \citet{provan16a} and ~\citet{provan18a} also determined both the amplitude ratio of the field oscillations in both systems, $k$, as well as the absolute amplitudes of the individual field components of each system.
The ratio of the amplitudes, \mbox{$k=B_{N0} / B_{S0}$}, where $B_{N\!,S0}$ are the amplitudes of the northern and southern PPO quasi-uniform `core' fields at radial distance $r < $12~\RS, respectively.
(1~\RS\ is Saturn's equatorial radius, 60268~km.)
The simultaneous presence of oscillations at two periods leads to `beating' of the two signals, affecting both the amplitude and phase of the combined signal.
Values of $k$ are determined through fits to beat-modulated magnetic field phase measurements within a given interval, and are plotted in Figure~\ref{fig:ts}d on a logarithmic scale.
With $k$ then fixed, the amplitudes of the separate northern and southern core fields $B_{N\!,S0}$ are then individually determined from fits to the magnetic field oscillations themselves.
A complete description of this procedure can be found in \citet{provan13a}.
Intervals with $k < 1$ correspond to southern system dominance, and vice-versa, while those periods with $k\approx1$ indicate approximately equal amplitudes of the northern and southern systems.
In principle, $k$ values less than $\sim$0.2 and greater than $\sim$5 ($\sim$$1/0.2$) should be taken as upper and lower limits, respectively, corresponding to intervals during which beat phase modulation arising from the weaker of the two systems is either undeterminable or absent.
Two such intervals dominated by the northern system do indeed occur following equinox, and are indicated by $k$ values shown by dashed lines.

A long-term trend is evident in the $k$ values shown in Figure~\ref{fig:ts}d, with $k$ consistently less than one in the early part of the Cassini mission through to late 2009, and $k$ consistently greater than one (where measured) from 2014 onwards, indicating an overall shift from southern dominance to northern dominance of the PPOs.
In the trans-equinox interval, beginning just prior to equinox but maintained for $\sim$4 years following this, large shifts in $k$ are observed about unity, with brief but clear periods of strong southern and northern dominance well after equinox, along with periods of apparently equal amplitudes in both systems.
This interval is the subject of specific analysis by~\citet{provan15a} and~\citet{fischer15a}.
This complex behaviour suggests that while seasonal variations may play some role in altering the relative amplitudes of the PPOs, for example through variations in illumination and therefore conductivity of the polar ionospheres, these cannot fully explain the observed properties of the system.

Core region amplitudes $B_{iS0}$ and $B_{iN0}$ are shown in Figure~\ref{fig:ts}e-g, where in each case the dashed red line shows the southern value $B_{iS0}$ and the blue solid line the northern value $B_{iN0}$, where $i$ is one of the three  spherical components of the field ($r$, $\theta$, $\varphi$, referenced to the planet's northern spin axis).
Azimuth $\varphi$ increases in the sense of planetary rotation.
These values are derived from fits to magnetic field amplitude data, using the determined phases and amplitude ratios $k$ shown in the upper panels.
These amplitudes, given in papers as indicated in Figure~\ref{fig:ts} are determined using field data obtained during the periapsis pass through the core region ($r < 12$~\RS) of each orbit of Cassini.
At all times, the ratio of the northern and southern amplitudes shown in Figures~\ref{fig:ts}e-g is therefore identical to the $k$ value shown in Figure~\ref{fig:ts}d, assumed equal for each component.
Similarly, $k$ is assumed equal for all components, and hence derived using a combination of the phase data from all three field components for a given interval.
Also indicated in Figures~\ref{fig:ts}e-g are averaged values $\overline{B}_{iS0}$ and $\overline{B}_{iN0}$ for each component, determined over all equatorial orbits studied in this paper.
These averaged amplitudes are later used for presentational purposes.

\section{Selection and processing of data}\label{sec:proc}

In this study we make use of 1-minute averaged field vectors from the Cassini fluxgate magnetometer~\citep{dougherty04a}.
The data are transformed into spherical polar `Kronographic' coordinate system, referenced to the planet's spin and magnetic axis, and processed initially by removing the internal (axisymmetric) field of the planet according to the model of~\citet{burton10a}.
At the typically large distances, $r>$3~\RS\ studied in this paper, our results are insensitive to the specific planetary field model used.
We then apply a band-pass filter to the data to suppress signals with periods outside the 5 to 20~h range encompassing the PPO, noting that the wide bandwidth retains and moreover does not distort the effects of Doppler-shifting induced by rapid azimuthal motion of the spacecraft at periapsis.
In a minor departure from~\citet{andrews10a}, we here use a so-called `zero-phase' filter, which we find performs marginally better in the vicinity of the very occasional data gaps than the Lanczos filter used previously.
Nevertheless, we apply the same restriction used by~\citet{andrews10a}, removing all otherwise useable data taken within 5~h of a data gap of duration greater than 5~h, both in order to prevent spurious filtered data distorting the analysis, and to provide commonality with that study.
Additionally, we remove data taken outside the magnetopause from further consideration, using a list of Cassini magnetopause crossings derived from that published by~\cite{pilkington15a}, updated by \cite{jackman19a} to span the remainder of the Cassini mission.
Cumulatively, the total data volume analysed in this study is $\sim$4.5 times larger than that of~\citet{andrews10a}.

\begin{figure}[htp]
\centering
\includegraphics[width=\textwidth]{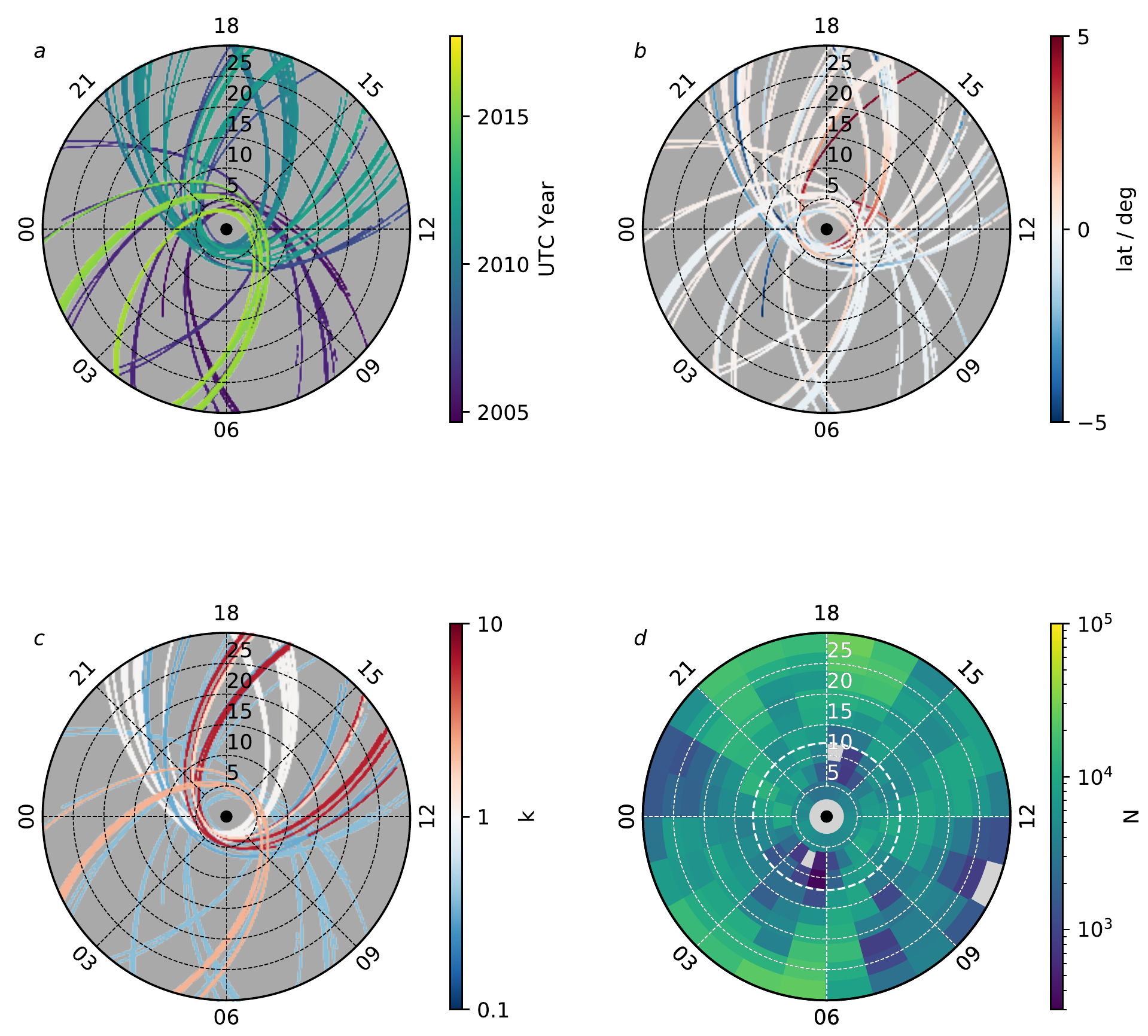}
\caption{Cassini's trajectory projected onto the equatorial plane.  Each panel shows the region out to $r=30$~\RS, with LT and radial distance values indicated. Cassini's trajectory color-coded according to a) time, b) latitude and c) the amplitude ratio $k$.  Panel d shows the number of 1~minute magnetic field vectors recorded within each spatial bin used in this study, as described in the text.
Bins with $N<300$ are shown grey.}
\label{fig:traj}
 \end{figure}

Only data obtained on orbits for which Cassini's peak absolute latitude was less than 5$^\circ$ were considered.
These orbits are highlighted in Figure~\ref{fig:ts} by the underplotted grey vertical bars.
The resultant coverage of the equatorial plane, indicated in Figure~\ref{fig:ts}c, is shown in full in Figure~\ref{fig:traj}, in which Cassini's position is projected onto the equatorial plane at 1~minute intervals and color-coded according to time, latitude, and amplitude ratio $k$ in panels a-c, respectively.
As can be seen in Figure~\ref{fig:traj}a, the sampling in time is not uniform owing to the varied orbital tour of Saturn.
Apoapses were located in the pre-midnight to post-dawn sector early in the mission, during which the southern PPOs were dominant, before the orbit plane was then rotated through the post-noon to post-dusk sector during the post-equinoctial second extended interval of near-equatorial orbits.
The final extended equatorial interval saw apoapsis return to the post-midnight to dawn sector during northern PPO dominant conditions, before the final high-inclination orbit series that terminated the mission.
Consequently, while the entire equatorial plane is now well-sampled across the whole mission, there nevertheless remain biases towards encountering certain LT sectors only during restricted intervals of time, e.g. the post-noon sector during the post-equinox interval, during which the northern and southern system periods approached and remained close.

As can be seen from Figure~\ref{fig:traj}b, virtually all of the data retained for the purposes of this study (86\%) were taken at latitudes $|\lambda| < 1^\circ$.
Of the remainder, 12\% were taken at southern latitudes and only 2\% at northern latitudes greater than 1$^\circ$.
However, while the vast majority (99\%) of the data were taken within $\pm$0.5~\RS\ of the spin-equatorial plane, the seasonal `warping' of Saturn's magnetodisc current sheet consequently affects the distribution of the data about the magnetospheric equator.
Using the model of~\citet{arridge08a}, we find that $\sim$30\% of the data used in this study were obtained northward of the current sheet center (taking a nominal `hinging' distance of 29~\RS).
This represents a marginal improvement in relative coverage of the two hemispheres compared with the interval studied by~\citet{andrews10a}.

In Figure~\ref{fig:traj}c, we color-code the trajectory according to the value of $k$, as shown in Figure~\ref{fig:ts}d, again noting that the sampling of the equatorial plane with respect to $k$ is not uniform.
There are, however, several extended regions which are sampled during intervals with both $k<1$ and $k>1$.
Nevertheless, the overall picture that emerges is one in which the data obtained throughout the equatorial plane is reasonably well distributed in both time and space, without any major biases that could lead towards conflating variations in one with the other.

Finally, in Figure~\ref{fig:traj}d we show the distribution of data used in this study, grouped into spatial bins, using an identical binning scheme to that employed by~\citet{andrews10a}.
Specifically, we use bins of 2~h width in LT, computed every hour and thereby overlapping the adjacent bin by 0.5~h.
Each non-overlapping radial bin is of 3~$\RS$ extent, with boundaries at 3, 6, \ldots 30~$\RS$.
The number of individual 1~minute magnetic field measurements, $N$, recorded in each bin is shown color-coded in Figure~\ref{fig:traj}d.
Those bins in which less than 300 samples (5~h of cumulative data) were obtained are shown grey, having been deemed not suitable for further analysis.
Half of the bins shown each contain at least $\sim$~100~h of cumulative magnetic field measurements.

\section{PPO field amplitude and phase determination}\label{sec:field}

Here we introduce the framework for the analysis of magnetic field oscillations used in this paper.
This is based primarily on that used in the previous study of equatorial PPO structures by~\citet{andrews10a}, updated to account for the simultaneous presence of northern and southern oscillations with differing rotation periods and amplitudes.

\begin{figure}[tp]
  \centering
  \includegraphics[width=8cm]{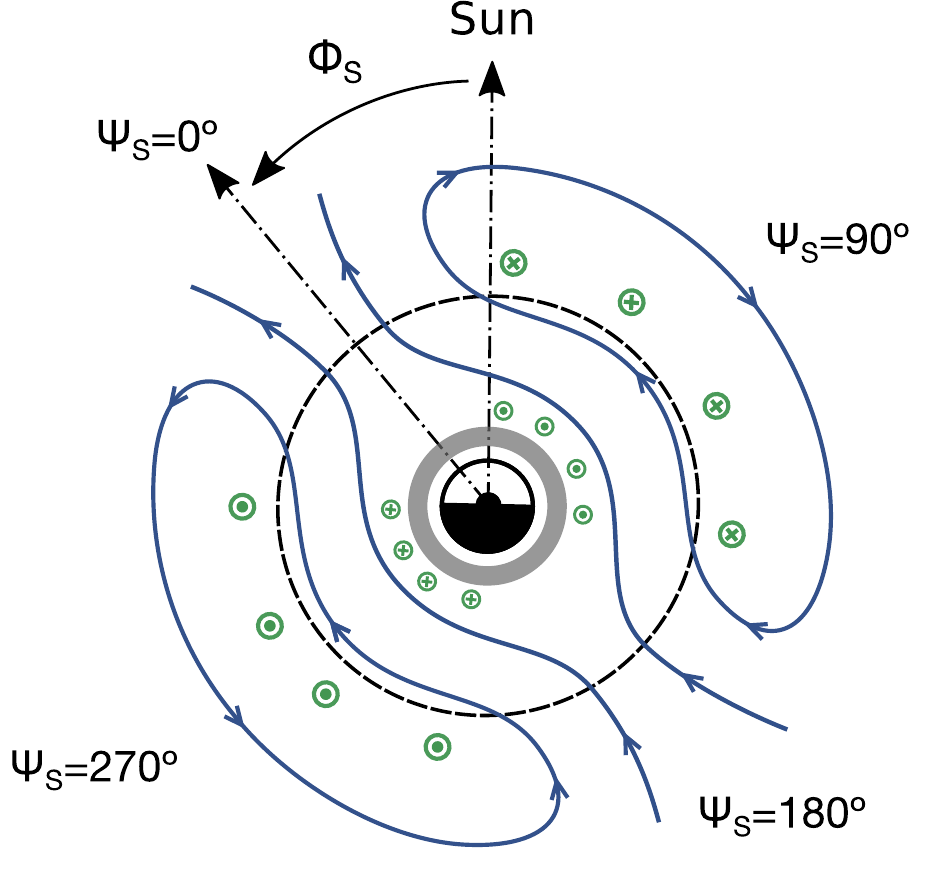}
  \caption{Simplified illustration (not to scale) of PPO magnetic field lines (blue lines) and field-aligned currents (green symbols) of the southern PPO system, after~\cite{andrews10a}.  The location of the Sun (noon meridian) is indicated, as is the principal meridian of the southern PPO phase coordinate system, $\Psi_S = 0^\circ$, offset from noon by $\Phi_S$.  The dashed circle indicates the approximate boundary of the core region, within which the PPO fields are quasi-uniform, and outside of which they take the form of a rotating transverse dipole.}
  \label{fig:ppo}
\end{figure}

We begin by defining the PPO phase of the northern (subscript N) or southern (S) systems,
\begin{equation}
  \Psi_{N\!,S}(t,\varphi) = \Phi_{N\!,S}(t) - \varphi,
  \label{eqn:psi}
\end{equation}
where the phase functions $\Phi_{N\!,S}(t)$ describes the instantaneous phase or orientation of the relevant system at time $t$.
Specifically, $\Phi_{N\!,S}$ defines the orientation of the principal $\Psi_{N\!,S}(t,\varphi)=0^\circ$ meridian of each system relative to the Sun, as is shown in Figure~\ref{fig:ppo}.
The quantity $\Psi_{N\!,S}$ then expresses the position of an observer at azimuth $\varphi$ in the rotating frame of the northern or southern PPOs, and serves to organise e.g.\ magnetic field perturbations, field-aligned currents, or other PPO-related phenomena.
The PPO phase $\Psi_{N\!,S}$ increases in the sense opposite to planetary rotation (anti-clockwise when viewed from the north).
Furthermore, for an observer moving in azimuth, Doppler-shifting of quantities fixed in the PPO frame is thereby accounted for in this expression.
The phase functions $\Phi_{N\!,S}(t)$ are themselves slightly non-linear functions of time, simply related to the PPO rotation periods $\tau_{N\!,S}$ through
\begin{equation}
  \tau_{N\!,S}(t) = 360^\circ \left(\frac{d \Phi_{N\!,S}(t)}{dt}\right)^{-1}.
  \label{eqn:tau}
\end{equation}
These rotation periods are shown in Figure~\ref{fig:ts}a.

A simplified schematic of the equatorial PPO transverse fields, currents, and associated plasma flows is also shown in Figure~\ref{fig:ppo}, adapted from a similar figure first shown by~\citet{andrews10a}.
The transverse component of the PPO field perturbation is shown by the blue lines.
The central reference for the study of PPO fields is taken to be the so-called `core' region field perturbations, those in the range \mbox{$3\leq r \leq 12$~\RS}, with the outer boundary indicated by the black dashed circle.
Within this region the field is quasi-uniform, with the azimuthal component $B_\varphi$ oscillating in lagging quadrature with the radial $B_r$ in both northern and southern systems, such that the field points along the $\Psi_{N\!,S}=0^\circ$ meridian.
Crucially, as discussed first by~\citet{andrews10b}, for the northern PPO system, the relative phase of the colatitudinal $B_\theta$ component is shifted by half a cycle in the core region as compared to the southern system, thus the colatitudinal $B_\theta$ component is in phase with $B_r$ in the southern system and in antiphase in the northern system.
Thus we have $B_r \approx \cos(\Psi_{N\!,S})$, $B_\varphi \approx \sin(\Psi_{N\!,S})$ and $B_\theta \approx \pm\cos(\Psi_{N\!,S})$, taking the $-$ sign term in the northern hemisphere system.
The presence of a non-zero $B_\theta$ perturbation is such that magnetic field lines associated with the quasi-uniform core field are tilted out of the equatorial plane, forming arches with apices towards the northern hemisphere for the southern system, and vice versa, though this is not shown in Figure~\ref{fig:ppo}.
The quasi-uniform core field perturbation is supported by a sinusoidally-varying current flowing through the equatorial plane at its outer boundary, indicated by the green icons outside the dashed black circle.
We typically refer to these as the `principal' PPO currents.
At larger radial distances, beyond the quasi-uniform core region, the phase of the azimuthal component is increased by 180$^\circ$ relative to the core, such that $B_\varphi\approx-\sin(\Psi_{N\!,S})$ and the field perturbation takes the form of a rotating transverse dipole oriented along the $\Psi_{N\!,S}=0^\circ$ meridian.
Corresponding currents flowing at the inner edge of the `core' region are also indicated in Figure~\ref{fig:ppo}, having opposite sign to the principal PPO currents flowing at larger distances, and acting to suppress the amplitude of the PPO fields towards the innermost core region.

The three components $i$ of the northern and southern PPO `core' equatorial magnetic fields are then defined as
\begin{equation}
  \label{eqn:bins}
  B_{iN\!,S}(\varphi, t) = B_{iN\!,S0}(t) \cos\left(\Psi_{N\!,S}(t, \varphi) - \gamma_{iN\!,S}\right),
\end{equation}
in which seasonal variations in amplitude are expressed through the time-varying amplitude $B_{iN\!,S0}(t)$ of each component.
The terms $\gamma_{iN\!,S}$ in equation~\eqref{eqn:bins} describe the constant phase of each core field component $i$ relative to the PPO phase $\Psi_{N\!,S}$.
Empirically these take the values $\gamma_{rS} = \gamma_{\theta S} = 0^\circ$ and $\gamma_{\varphi S} = 90^\circ$ for the southern PPO system and $\gamma_{rN} = 0^\circ$, $\gamma_{\theta N} = 180^\circ$ and $\gamma_{\varphi N} = 90^\circ$ for the northern.
Thus, for both the northern and southern systems, the radial component of the PPO core field reaches its maximum positive value where $\Psi_{S}(t,\varphi)=0^\circ$.

The total combined core field in component $i$ is then described simply by the sum of $B_{iS}(\varphi, t)$ and $B_{iN}(\varphi, t)$.
When both northern and southern system PPOs are present with comparable amplitudes, the resultant magnetic field oscillations will exhibit constructive and destructive interference according to the beat phase $\Delta\Phi_{NS}(t) = \Phi_N(t) - \Phi_S(t)$.
The differing field structures are such that when constructive interference takes place between the radial and azimuthal components at $\Delta\Phi_{NS}=0^\circ$, destructive interference is present in the colatitudinal component, with the situation reversed at $\Delta\Phi_{NS}=180^\circ$ (modulo-360$^\circ$).

\section{Equatorial PPO field structure}\label{sec:eq}
\subsection{Independent determination of northern and southern equatorial field structures}\label{sub:indep}

\citet{andrews10a} determined the amplitude and phase structure of the equatorial PPO fields as a function of position, extending the analysis beyond the `core' region, and considering variations with radial distance and LT.
However, their analysis was undertaken before the presence of independent northern and southern PPO systems was fully understood, and as such took no account of the influence of the (then much weaker) northern system on the dominant southern system.
Expressed in the above scheme,~\citet{andrews10a} determined spatially varying amplitudes $B_{iS0}(r,\varphi)$ and phases $\psi_{iS}(r,\varphi)$ by fitting the expression
\begin{equation}
  B_{iS}(r,\varphi,t) = B_{iS0}(r,\varphi) \cos\left(\Psi_{S}(t) - \psi_{iS}(r,\varphi) \right),
  \label{eqn:a10fit}
\end{equation}
to each component $i$ of the processed magnetic field using data grouped into spatial bins.
The amplitudes $B_{iS0}(r,\varphi)$ and phases $\psi_{iS}(r,\varphi)$ were taken as free parameters and fitted to the data, their values taken to be constant within each spatial bin, and furthermore constant in time across the interval studied.
Equatorial magnetic field data used by~\citet{andrews10a} were obtained from August 2004 to December 2007 (Revs A-54 - see Figure~\ref{fig:ts}), thus reflective of an extended interval of apparently stable southern-system dominance ($k\approx $0.4).

Here we extend the analysis of~\citet{andrews10a}, by considering the simultaneous presence in the data of oscillations associated with independently rotating northern and southern systems.
A first approach to this can be taken by applying the same `single hemisphere' approach to the processing of the data as used by~\citet{andrews10a}, repeating the analysis for each hemisphere independently with the substitution of the appropriate phase function $\Phi_{N}(t)$ or $\Phi_{S}(t)$ into equation~\eqref{eqn:a10fit}.
Given sufficient data coverage and distinct and well-determined rotation periods $\tau_N$ and $\tau_S$, analysing the data with respect to one hemisphere is feasible even under the presence of a significant modulation due to the opposing hemisphere, because the two phases (modulo 360$^\circ$) are overall uncorrelated.
The effect of the field oscillations associated with the `opposing' hemisphere is merely to add noise to the data, which rapidly sums towards zero excepting  those intervals where their respective rotation periods are close (or indeed, equal to within measurement uncertainties, such as was found e.g.\ by~\citet{provan16a} in the interval from mid 2013 to mid 2014).

Cassini's orbital tour is such that it cumulatively samples most of the equatorial plane over the whole mission (Figure~\ref{fig:traj}).
However, given that the core region amplitudes $B_{iN\!,S0}$ also vary during the course of the mission as shown in Figure~\ref{fig:ts}e-g, the function fitted to equatorial field data must be modified to account for these seasonal variations.
We therefore instead fit the expression
\begin{equation}
  \label{eqn:bins_new}
  B_{iN\!,S}(r, \varphi, t) =  f_{iN\!,S}(r, \varphi) B_{iN\!,S0}(t)\cos ( \Psi_{N\!,S}(t,\varphi) - \gamma_{iN\!,S} - \xi_{iN\!,S}(r,\varphi) ),
\end{equation}
to equatorial field data, treating each hemisphere (N or S) entirely independently, using the core region amplitudes $B_{iN\!,S0}$, phases $\Psi_{N\!,S}$ and relative phases $\gamma_{iN\!,S}$ introduced in section~\ref{sec:field}.
The terms $f_{iN\!,S}$ and $\xi_{iN\!,S}$ are then free parameters that describe the amplitude and phase, respectively, of the measured field oscillations in each bin relative to the core region, and are then obtained by fitting the data using a non-linear least-squares algorithm (a standard Levenberg-Marquardt algorithm).
For each of the northern and southern systems equation~\eqref{eqn:bins_new} is thus a function only of quantities relevant to that hemisphere.

\begin{figure}[htp]
  \includegraphics[width=\textwidth]{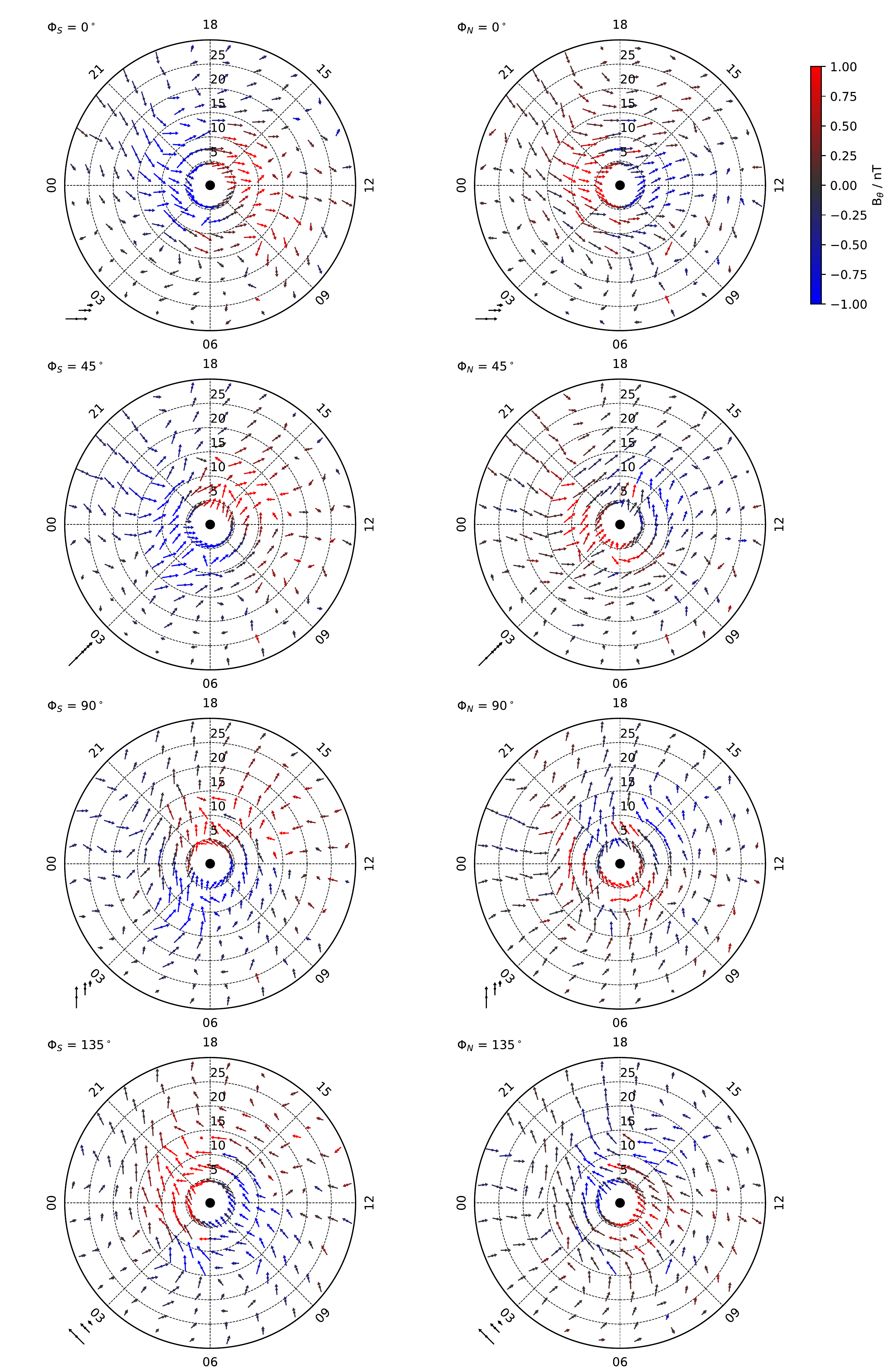}
  \caption{Caption next page.}
\end{figure}
\addtocounter{figure}{-1}
\begin{figure}
  \caption{Plots showing the equatorial magnetic field oscillations derived independently for the northern and southern PPO systems from fits to data of equation~\eqref{eqn:bins_new}.
  Field vectors are computed at phases $\Phi_{N\!,S}=0^\circ$, 45$^\circ$, 90$^\circ$, and 135$^\circ$, and using the overall averaged core amplitudes shown by the dotted red (S) and blue (N) lines in Figure~\ref{fig:ts}e-g.
  The left panels a-d show field vectors determined from fits to southern system phases and amplitudes, while the right panels e-h show northern equivalents.
  Each panel depicts the transverse component ($r$-$\varphi$) within each bin by an arrow along the field direction at the phase $\Phi_{N\!,S}$ indicated.
  The arrows are non-linearly scaled, with the icons in the lower-left of each panel indicating 0.1, 1, and 5~nT vectors oriented along the expected instantaneous direction of the transverse core region field.
  The colatitudinal component is color-coded using the color table, such that positive values, shown by red colors, are into the plane of the page (and negative blue outward).
  Each panel is labelled according to the oscillation phase $\Phi_{N\!,S}$, and is spaced by $45^\circ$ from the previous.
  The full 360$^\circ$ cycle of $\Phi_{N\!,S}$ can be obtained by reversing both the sense and color of each vector (inverting the colour table used to display the $B_\theta$ component), and adding 180$^\circ$ to the indicated phase of each panel.}
  \label{fig:indep_ns}
\end{figure}

In Figure~\ref{fig:indep_ns} we show results of this fitting procedure as applied to magnetic field measurements, processed and binned spatially as described in section~\ref{sec:proc}.
Results are presented in a format similar to that used previously by~\citet{andrews10a}, in which field vectors are plotted at the center of each spatial bin using the relative amplitudes and phases determined from the fits, at a given specified rotation phase $\Phi_{N\!,S}$.
The left column of panels shows fits of equation~\eqref{eqn:bins_new} taking only the southern hemisphere terms, while the right column of panels shows the corresponding northern hemisphere terms, again stressing that the fits are performed entirely independently of one another, with no additional prior processing of the data.
Within each bin, having determined both $f_{iN\!,S}$ and $\xi_{iN\!,S}$ from the field data, field vectors are reconstructed using the averaged values of the core region amplitudes $\overline{B}_{iN\!,S0}$, as given in the second column of Table~\ref{tab:stuff} and shown in Figure~\ref{fig:ts}.
The averaged amplitudes $\overline{B}_{iN\!,S0}$ are computed from the piecewise determined instantaneous values of the core region amplitudes in each field component only over those intervals for which equatorial data were obtained, i.e.\ those intervals shaded grey in Figure~\ref{fig:ts}.
(Remaining quantities given in Table~\ref{tab:stuff} are introduced in the following sections.)
We note that the components of the averaged core region amplitudes are almost equal in the opposing hemispheres, i.e.\ $\overline{B}_{iN0}\approx\overline{B}_{iS0}$, and furthermore close to those measured during the interval of $k\approx1$ conditions immediately post-equinox.
The ensemble of vectors shown in Figure~\ref{fig:indep_ns} thus strictly represent mission-averaged results, whilst also being reasonably close to those that may be expected post-equinox.

The individual panels a-d shown in the left column of Figure~\ref{fig:indep_ns} show the evolution of the southern PPO equatorial field through one half of the full cycle, from $\Phi_S=0^\circ$ to 135$^\circ$.
The remaining half-cycle can be visualised from these panels by simply reversing the sense and color of the plotted vectors and adding 180$^\circ$ to the value of $\Phi_S$ corresponding to each panel.
Panels e-h in the right column of Figure~\ref{fig:indep_ns} show corresponding northern PPO equatorial fields in the same format.
Without discussing the detailed structure of the PPO fields plotted at this stage, the commonality between the northern and southern systems is nevertheless readily apparent when comparing each pair of panels.
As noted previously, the transverse components of the oscillations are expected to be in phase with one another, at least in the core region, when $\Phi_N = \Phi_S$.
Comparing each pair of panels in Figure~\ref{fig:indep_ns}, this can be seen to be the case, with each individual plotted vector generally aligned with its corresponding icon in the opposite hemisphere, to a good approximation, and furthermore having similar amplitude.
This correspondence is generally seen to be best where the amplitudes are themselves largest, i.e. within the core region and into the nightside, and is weaker on the dayside where amplitudes are typically smaller.
Meanwhile, the sense of the colatitudinal component, depicted by the color of the individual vectors, is expected to be opposite in the two hemispheres when $\Phi_N = \Phi_S$, owing to the antiphase relationship of these components in the opposing hemispheres.
Again, this expectation is realised to a large degree when comparing the pairs of panels in Figure~\ref{fig:indep_ns}, with those vectors shown red (positive) in the southern hemisphere panels typically shown blue (negative) at the same location in the northern hemisphere.
However, the correspondence between the field vectors displayed in each pair of panels is not perfect.
Discrepancies are clear in certain individual bins, particularly towards larger radial distances on the dayside.
Despite this, no evidence is seen in these plots of any systematic difference between the spatial structure of the PPO fields in the two hemispheres.
Where differences are apparent, these are constrained to the extent of individual bins and are generally not reflected even in immediately adjacent bins.

\begin{figure}[tp]
  \includegraphics[width=\textwidth]{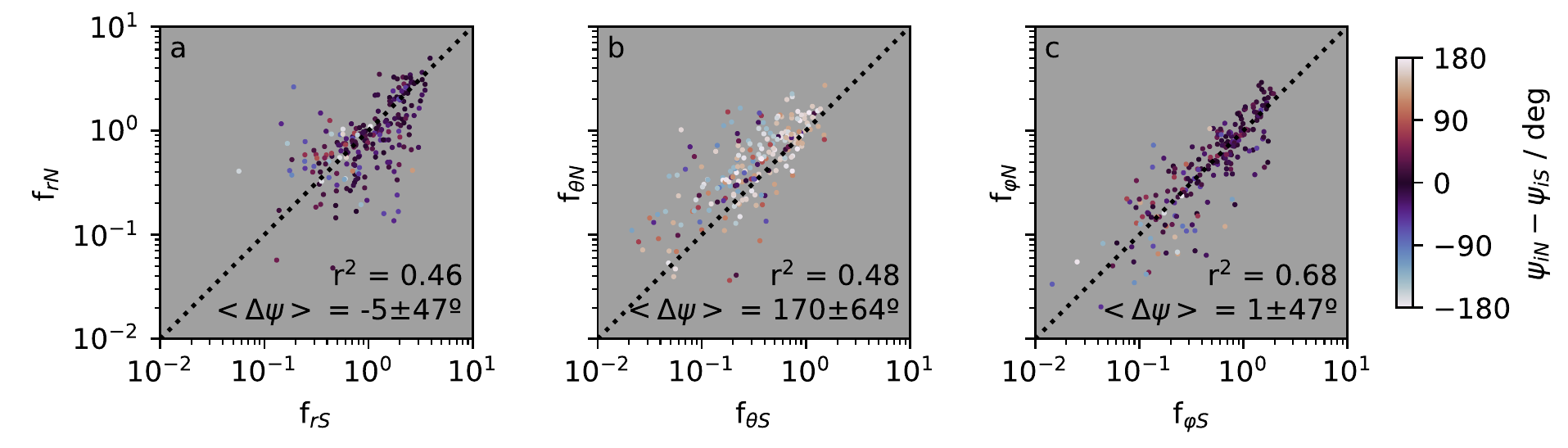}
  \caption{Comparison of independently determined northern and southern relative amplitudes and phases for the three field components within each spatial bin used in this study.  Relative amplitudes $f_{iN\!,S}$ and phases differences $\Delta\psi = \psi_{iN} - \psi_{iS}$ are depicted for each component using the axes and the color-code, respectively.}
  \label{fig:ns}
\end{figure}

\begin{table}[tp]
\begin{tabular}{cc|cccccc}
  \hline
Hemisphere & Component $i$ & $\gamma_{iN\!,S}$& $\overline{B}_{iN\!,S0}$ & $\overline{\xi}_i$, core & $\overline{f}_i$, core & $\overline{\psi}_{iN\!,S}$, core & $\overline{v}_{ri,N\!,S}$  \\
& & [deg] & [nT] & [deg] & & [deg] & [km s$^{-1}$ $(^\circ/ \mathrm{R_S}$)] \\\hline
    & $r$       & 0   & 0.68 & -3.8$\pm$21 &  1.01$\pm$0.48 & -3.8$\pm$21 & 384 (1.47) \\
N   & $\theta$  & 0   & 1.13 & -9.6$\pm$22 & 1.03$\pm$0.29 & 170.4$\pm$22 & 123 (4.60) \\
    & $\varphi$ & 90  & 1.21 & -4.6$\pm$17 & 0.85$\pm$0.30 & 85.4$\pm$17 & 197 (2.85) \\ \hline
    & $r$       & 0   & 0.66 &  &  & -3.8$\pm$21 &\\
S   & $\theta$  & 180 & 1.17 & (as N) & (as N) & -9.6$\pm$22 & (as N)\\
    & $\varphi$ & 90  & 1.29 &  &  & 85.4$\pm$17  &\\  \hline

\end{tabular}
\caption{Parameters referenced in the text.  Amplitudes and phases given are calculated in the core region, $r<15$~\RS.  Radial phase velocities are calculated using all available data, although excluding the azimuthal component at radial distances greater than 15~\RS.}
\label{tab:stuff}
\end{table}

We further quantify the correspondence in the spatial variations of amplitude and phase between the northern and southern systems in Figure~\ref{fig:ns}.
For each component $i$ in Figure~\ref{fig:ns}a-c, respectively, we directly compare relative amplitudes determined independently for the northern and southern systems, by plotting $f_{iN}$ verus $f_{iS}$.
Each plotted point thus corresponds to a comparison between fits in the same spatial bin, to the self-same field data, using the northern and southern forms of equation~\eqref{eqn:bins_new}.
Values lying on the dotted line of unit slope thus indicate equal relative amplitudes determined at a particular location.
Corresponding values of the $R^2$ statistic are given in each plot.
Agreement is seen to be best in the case of the azimuthal component, having the highest $R^2$ statistic, likely due to the generally lower `background' present in this component (neither the internal planetary field nor the azimuthal ring current field contributing to this component).

The difference between fitted phases $\Delta\xi_{i}=\xi_{iN} - \xi_{iS}$ in each bin are shown by the color coding of each plotted point in Figure~\ref{fig:ns}.
Directional means and standard deviations of these plotted phase differences are given in the figure, and are typically close to zero in the radial and azimuthal components, while being close to 180$^\circ$ in the colatitudinal component, as expected (to within the approximately quarter-cycle spread in phase difference, i.e. plus or minus one standard deviation).
Corresponding standard errors on the given mean values are less than 5$^\circ$ in all components.
Thus, given the available data, no convincing evidence is found for any major difference between the spatial phase and amplitude structures of the northern and southern PPO systems.

\subsection{Determination of common field structure}\label{sub:common}

The analysis presented in section~\ref{sub:indep} suggests no systematic differences in the spatial structure of the amplitude and phase of PPO fields associated with the northern and southern systems.
Taking this to be the case, a more accurate examination of the spatial structure common to both systems can be obtained using a single fit to the data in each bin, simultaneously determining the common deviation in phase and amplitude from both PPO core regions.
This approach therefore directly accounts for the beat-modulation of both systems, rather than assuming that the influence of this effect averages to zero over the data set as a whole.

Here we again define both amplitudes $f_i(r,\varphi)$ and phases $\xi_i(r,\varphi)$ for each field component $i$ relative to the core region, but now take them to be common to both northern and southern PPO systems (and therefore drop the subscript $N$ or $S$ from both terms).
The instantaneous combined field oscillations at the location of the spacecraft can then be written as
\begin{subequations}
  \begin{eqnarray}
    \label{eqn:commonfit}
    B_i(r, \varphi, t) =  f_i(r, \varphi) & \Big[ & B_{iS0}(t)\cos ( \Psi_S(t,\varphi) - \gamma_{iS} - \xi_{i}(r,\varphi) ) \nonumber \\
     && + B_{iN0}(t)\cos ( \Psi_N(t,\varphi) -  \gamma_{iN} - \xi_{i}(r,\varphi) ) \Big]
  \end{eqnarray}
or equivalently,
  \begin{equation}
    B_i(r, \varphi, t) =  f_i(r, \varphi)  B_{iS0}(t) \Big[  \cos ( \Psi_S(t,\varphi) - \gamma_{iS} - \xi_{i}(r,\varphi) ) + k(t) \cos ( \Psi_N(t,\varphi) - \gamma_{iN} - \xi_{i}(r,\varphi) ) \Big],
  \end{equation}
\end{subequations}
where each component reaches its maximum and minimum values where \mbox{$\Psi_{N\!,S} = \gamma_{iN\!,S} + \xi_i$} and \mbox{$\Psi_{N\!,S}=\gamma_{iN\!,S}+\xi_i+180^\circ$} (modulo-360$^\circ$), respectively.
All necessary input parameters to the fit over the interval studied here, i.e., $B_{iN\!,S0}(t)$, $\Psi_{N\!,S}(t)$ and $\gamma_{iN\!,S}$,   have been determined by~\citet{andrews12a} and \citet{provan13a, provan16a, provan18a}, as summarised in Figure~\ref{fig:ts}.
We note also that the magnetic field data within a particular bin need not be continuous in time, with a minimum of one orbital period elapsing before the spacecraft returns to the same location in the equatorial plane.

\begin{figure}[htp]
  \includegraphics[width=\textwidth]{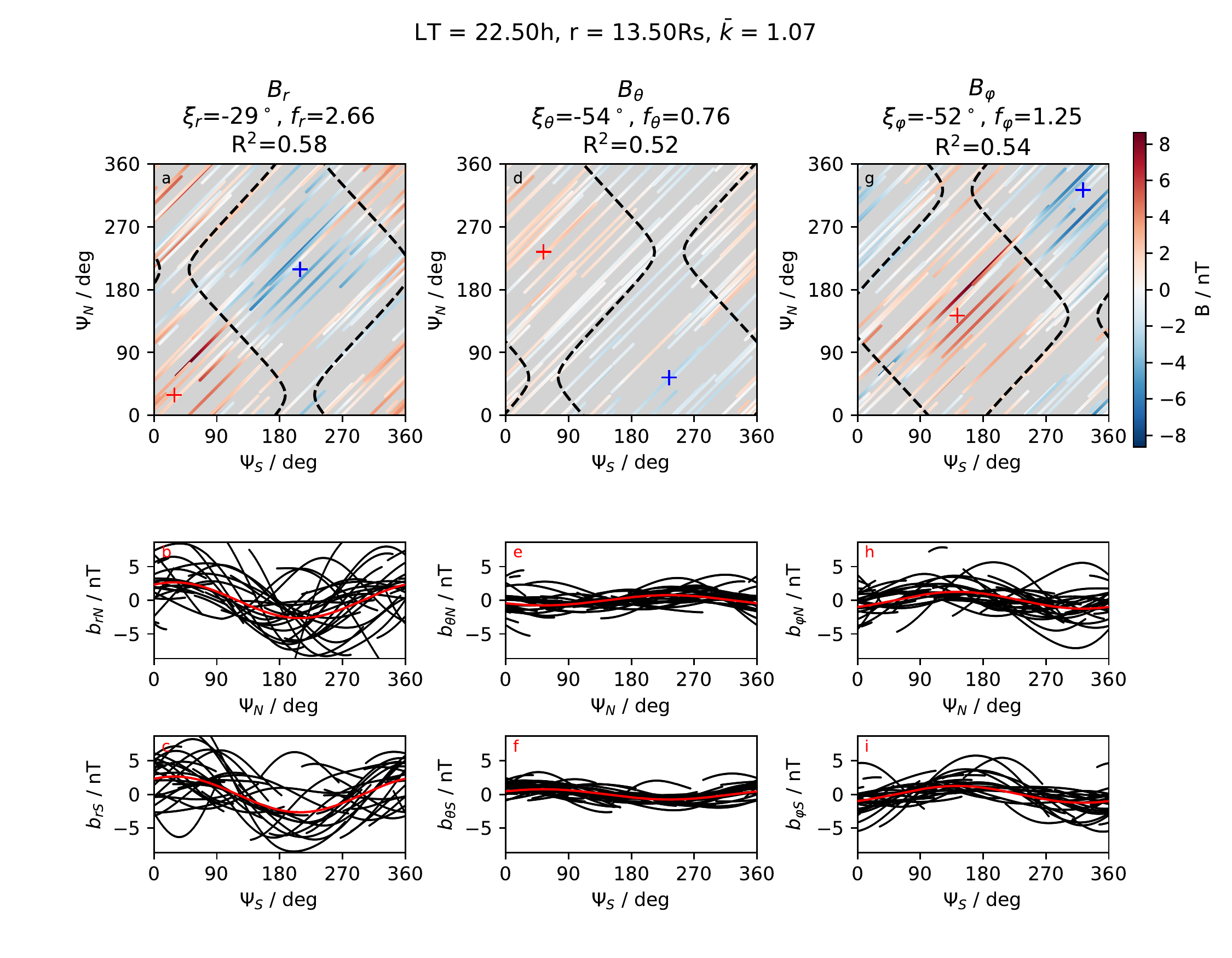}
  \caption{An example of fits of equation~\eqref{eqn:commonfit} to magnetic field data obtained in the equatorial plane in the range of LT from 21:30 to 23:30~h and radial distance $r$ from 12 to 15~\RS.
  Panel a shows processed radial magnetic field data $B_r$(t) colour-coded versus southern and northern phase $\Psi_S$(t) and $\Psi_N$(t) (modulo-360$^\circ$).
  The dashed lines through the panel separate the positive and negative parts of the combined oscillation cycles, as determined from the fit of equation~\eqref{eqn:commonfit}.
  The time-averaged value of $k$ computed for the data shown, $\overline{k}$, is noted at the top of the panel, along with the determined values of $f_i$, $\xi_i$, and the coefficient of determination $R^2$ for each component $i$.
  Panel b shows scaled magnetic field values \mbox{$b_{rN} = B_{r}(t) / B_{rN0}(t)$} (black traces) versus northern phase $\Psi_N(t)$.
  The quantity \mbox{$f_r \cos ( \Psi_N(t,\varphi) -  \gamma_{rN} - \xi_{r})$} is overplotted as the red line, with values $f_r$ and $\xi_r$ determined from least-squares two-phase fits to all magnetic field data in this bin, as discussed in the text.
  Panel c is equivalent to panel b, but shows instead southern-system scaled fields and fit, plotted versus southern phase $\Psi_S(t)$, and overplotted \mbox{$f_r \cos ( \Psi_S(t,\varphi) - \gamma_{rS} - \xi_{r})$}.
  The remaining panels d - i are identical to their adjacent panels a - c, but showing field values and fits for the remaining two components, $B_\theta$ and $B_\varphi$ as indicated.
  }
  \label{fig:fit}

\end{figure}

By way of example, we show in Figure~\ref{fig:fit} a fit of equation~\eqref{eqn:commonfit} to data obtained in the equatorial region defined by \mbox{$12\;\RS\ < r < 15\;\RS$} and 21:30~h~\textless~LT~\textless~23:30~h, processed as outlined in section~\ref{sec:proc}.
This bin contains $\sim$210~h of magnetic field measurements, cumulatively, representing 12750 individual field vectors.
The majority, $\sim$75\%, of these measurements were taken during 2009-2011, with the remainder obtained during brief intervals in 2006 and 2015.
The averaged ratio of the northern to southern core region amplitudes $\overline{k}$ over this sub-set of the data is 1.07, thus indicative of an overall weak dominance of the northern system over the southern.
Each column within the figure corresponds to one of the field components $i$.
The uppermost panels, Figures~\ref{fig:fit}a, d and g, show the three components $B_r$, $B_\theta$ and $B_\varphi$ of the magnetic field colour-coded versus both total northern and southern phases, $\Psi_N$ and $\Psi_S$.
Equation~\eqref{eqn:commonfit} is fitted to these data using the non-linear least squares algorithm mentioned in~section~\ref{sub:indep}.
For the data obtained within this spatial bin, organisation of the magnetic field by both phase functions is apparent in each field component, as evidenced by the `grouping' of the positive and negative values towards unique phase values in both coordinates.
The phases of the maxima and minima in the field are indicated by the red and blue crosses in the upper panels, i.e.\ the points where  \mbox{$\Psi_{N\!,S}(t,\varphi) = \gamma_{iN\!,S} + \xi_{i}(r,\varphi)$} and \mbox{$\Psi_{N\!,S}(t,\varphi) = \gamma_{iN\!,S} + \xi_{i}(r,\varphi)+180^\circ$}, respectively.
Black dashed lines show the zeroes of the fitted equation~\eqref{eqn:commonfit}, as determined from the respective values of $\xi_i$ for each component, using the average value of the amplitude ratio $\overline{k}$ determined over the interval of data included in this bin.

The lower two rows of panels show, for the northern and southern systems independently, the same magnetic field measurements, scaled and plotted versus the respective phase functions.
Specifically, the central row of panels Figures~\ref{fig:fit}b, e and h, shows as black traces scaled field values \mbox{$b_{iN} = B_{i} / B_{iN0}(t)$} for each component $i = r$, $\theta$ and $\varphi$, plotted versus the northern phase function $\Psi_N(t,\varphi)$ .
Data are shown scaled according to $B_{iN0}$ in order to remove temporal variations in the amplitude, which may well be significant within a single bin, Cassini's orbital tour often returning it to the same location as part of a different orbit sequence spaced widely in time.
The overplotted red traces in Figures~\ref{fig:fit}b, e and h then show the quantity \mbox{$f_i\cos(\Psi_N(t,\varphi) - \gamma_{iN} - \xi_{i})$}, with $f_i$ and $\xi_i$ as determined by the fit to the field data, and thereby indicating the variation relative to the northern system core PPO oscillation.

The bottom row, Figures~\ref{fig:fit}c, f and i, then show the same field data scaled instead by the southern system amplitudes $B_{iS0}(t)$ and plotted versus southern phase $\Psi_S(t,\varphi)$.
The overplotted quantity is then \mbox{$f_i\cos(\Psi_S(t,\varphi) -  \gamma_{iS} - \xi_{i})$}, with identical values of $f_i$ and $\xi_i$ to that in each of the panels above, respectively.
Consequently, the red overplotted lines are entirely identical in Figures~\ref{fig:fit}b and c, and in Figures~\ref{fig:fit}h and i, while they are shifted by 180$^\circ$ of phase relative to each other in Figures~\ref{fig:fit}e and f (as $\gamma_{rN} = \gamma_{rS}$ and $\gamma_{\varphi N} = \gamma_{\varphi S}$, while $\gamma_{\theta N} = \gamma_{
\theta S} + 180^\circ$).
We note again that these red lines in the lower six panels of Figure~\ref{fig:fit} do not represent fits to the underplotted data  themselves, but rather indicate the level of variation that can be ascribed to the separate northern and southern system oscillations.

Figure~\ref{fig:fit} illustrates the degree of success of this simple fitting scheme in describing the data.
Looking at the upper row of panels, the data are seen to be well-grouped simultaneously in both northern and southern phase, as expected for data taken with an average $k$ only slightly above unity, indicating approximately equal northern and southern system amplitudes in each component.
The dashed lines in these panels, corresponding to the zeros of the combined (northern and southern) field oscillations do indeed broadly separate positive and negative displacements of the measured fields from zero, and the locations in phase of the maximum and minimum amplitudes are also in each case commensurate with the underlying data.
Coefficients of determination (R$^2$) are given in the plot for each of the fits, and indicate that $\sim$50-60\% of the variance in the processed field data is accounted for by this model.
The remaining `unexplained' variance is principally attributable to uncertainties both in the determined rotation periods of the two systems, the core region amplitudes $B_{iN\!,S0}$, and temporal variations in these quantities occurring on shorter timescales than those over which they are determined (c.f., Figure~\ref{fig:ts}).
Factors external to the PPO fields, such as variations in the various quasi-static magnetospheric current systems associated with changes in the solar wind conditions also likely contribute to the scatter in the data about the fitted values, with this effect in particular likely to be the dominant source of error at larger radial distances.

Turning now to the lower panels of Figure~\ref{fig:fit}, it can be appreciated that during the majority of the interval for which these data are obtained, no single hemisphere is dominant over the other.
This is evident as, for each component, significant scatter is present about the system-average oscillation shown by the red line, with the majority of this scatter being due to the presence of oscillations of comparable amplitudes associated with the opposing hemisphere.
In this instance, the oscillations can be seen to be markedly enhanced relative to the core in the radial component, with scaled field amplitudes $b_{rS/N}$ significantly larger than unity and $f_r=2.66$.
This enhancement is also present, to a lesser extent, in the azimuthal component fields, $f_\varphi=1.25$, while the colatitudinal component is instead marginally suppressed, $f_\theta=0.76$.

\subsection{Common amplitude and phase structure}\label{sub:str}
We now apply the fitting procedure introduced in section~\ref{sub:common} to all equatorial field data, binned using the scheme described in section~\ref{sec:proc}.
We proceed by first showing the determined values of the relative amplitudes $f_i(r,\varphi)$ and phases $\xi_i(r,\varphi)$.
From these, phase velocity profiles are then computed, and following this, equatorial field vectors are computed.

\begin{figure}[htp]
  \includegraphics[width=\textwidth]{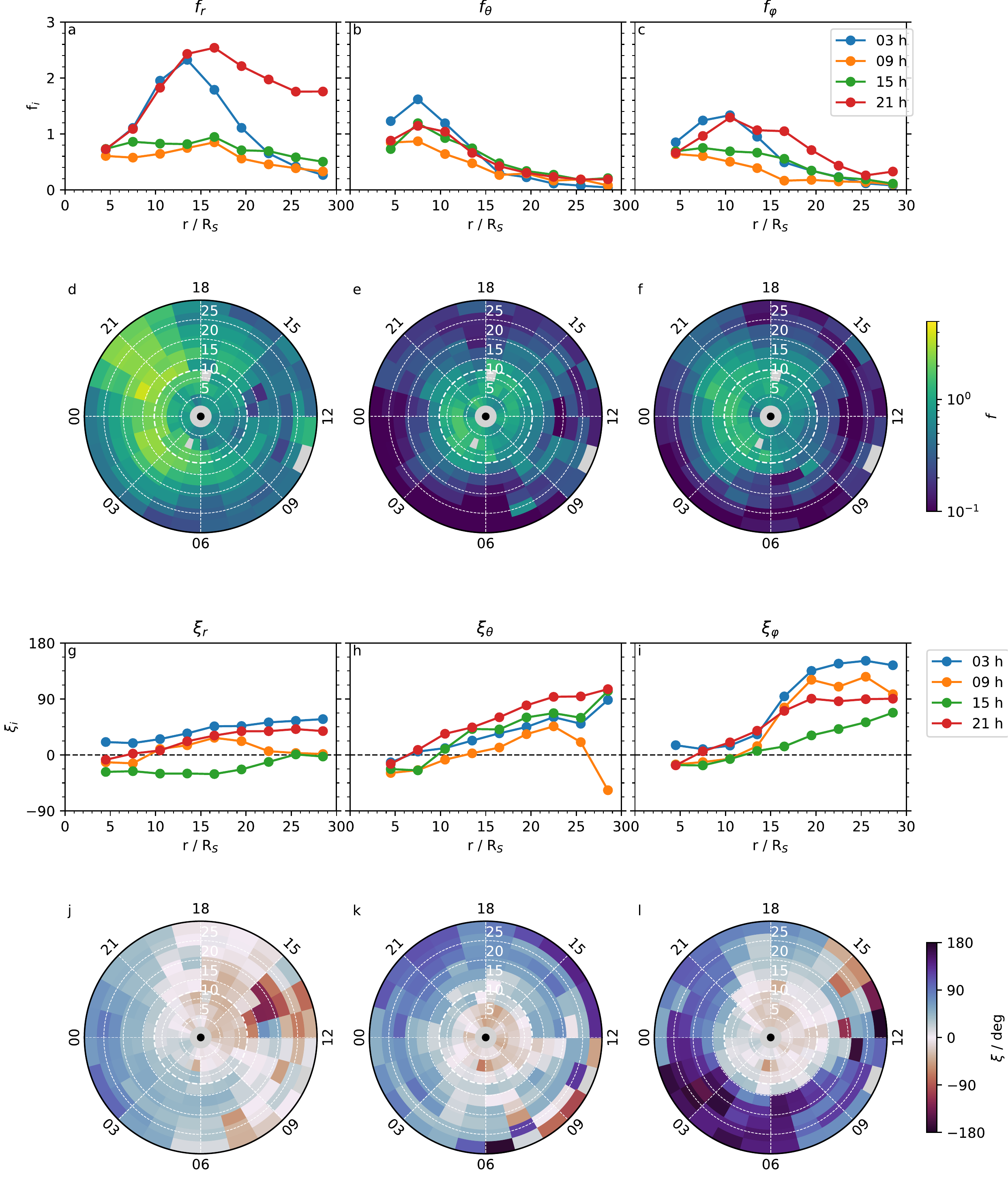}
\caption{Plots showing the amplitude and phase of the equatorial PPOs relative to the core region, as determined in spatial bins throughout the equatorial plane.
  Panels a-c show relative amplitudes $f_i$ for each component $r$, $\theta$, and $\varphi$, respectively, versus radial distance averaged over LT bins in four quadrants of LT centered on 03, 09, 15 and 21~h (colors indicated in the legend).
  Panels d-f show color-coded values of $f_i$ throughout the equatorial plane.
  Radial distance and LT in units of $\RS$ and hours are indicated in each panel, as is the outer boundary of the core region at 12~\RS\ (dashed white line).
  Panels g-i and j-l show relative phases $\xi_{i}$ versus radial distance at fixed LT, and throughout the equatorial plane in a format corresponding to those of the first six panels.
  }
  \label{fig:str}
\end{figure}

Results of this fitting procedure are shown in Figure~\ref{fig:str}.
Each column displays results determined for one of the three field components, as indicated.
The upper six panels show relative amplitudes $f_i$, while relative phases $\xi_i$ are displayed in the lower six.
Figures~\ref{fig:str}a-c show the variation of relative amplitudes $f_i$ versus radial distance in four separate quadrants of LT centered on 03, 09, 15 and 21~h, serving to indicate large-scale variations of the PPO fields relative to the core.
Values shown in each quadrant are determined by combining data from all spatial bins shown in Figure~\ref{fig:traj}d that lie within this quadrant at a given radial distance, thus indicating the large-scale structure of the PPO fields.
Figures~\ref{fig:str}d-f then show $f_i$ for each component, color-coded throughout the equatorial plane, each colored sector corresponding to a single bin depicted in Figure~\ref{fig:traj}d.
For both sets of panels a-c and d-f, a value near to unity indicates a location where the measured PPO amplitude is close to that determined in the core region, therefore following the secular variations in the (superposed) system amplitudes $B_{iN0}(t)$ and $B_{iS0}(t)$ during the interval studied.
Meanwhile, values above or below unity indicate a region in which the observed PPO field amplitudes are on average above or below the respective system amplitudes.
Those few bins at large radial distances, beyond $\sim$25~\RS\ on the dayside which exhibit significantly larger values of $f_i$ than their immediate neighbours are likely the result of poor fits to the data, the result of the increased effect here of field perturbations associated with variations in the solar wind conditions.
Similarly, Figures~\ref{fig:str}g-i and j-l and show the relative phase $\xi_i$, for each component, first as a function of radial distance in four LT quadrants in the same format as panels a-c, and then color-coded throughout the equatorial plane in the same format as panels d-f.
Values of $\xi_i$ are shown colour-coded using a cyclical color table appropriate to these angular data, determined to modulo-360$^\circ$.
We stress that the data shown in Figure~\ref{fig:str} are not themselves specific to either the northern or southern system, but instead show how the amplitude of the PPO magnetic field is locally modified compared to mean core region amplitudes $B_{iN0}(t)$ and $B_{iS0}(t)$ (and likewise the phases).

Examining first the relative amplitudes of the colatitudinal component $f_\theta$, shown in Figure~\ref{fig:str}b and e, it can be seen that values are maximal inside the core region, both in the four indicative quadrants of LT plotted in the upper panel, and indeed elsewhere as displayed immediately below.
Taking a simple average over all bins inside 12~\RS, we find a core-region mean value of $\overline{f}_\theta=1.03\pm0.29$, i.e.\ close to the unit value expected, and where the quoted uncertainty of 1-standard deviation represents the scatter in values within individual bins.
(These and other values derived in this section are given in Table~\ref{tab:stuff}.)
Beyond the core region, $f_\theta$ decreases to $\sim$10-20\% of the core region value at the outer boundary of the region studied.
Minor variations in $f_\theta$ with LT are evident in the core region.
A larger and more systematic variation in $f_\theta$ is present outside the core region, where values appear to be systematically lower in a wide band of LT centered on $\sim$03~h outside the core region.
Data obtained in this sector were taken during two distinct intervals of the mission (c.f.~Figure~\ref{fig:traj}a), and therefore the generally lower relative amplitudes found here are unlikely the result of restricted sampling in time.
However, variations in $f_\theta$ with LT are generally less significant than those seen in the remaining components.

Turning to the relative amplitudes of the radial component, $f_r$ shown in Figure~\ref{fig:str}a and d, a rather different situation is apparent.
While the mean value $\overline{f}_r=1.01\pm0.48$ in the core region is again in line with expectations, maximal values of $f_r$ are reached well outside of the core region in the two nightside quadrants of LT shown in Figure~\ref{fig:str}a.
Indeed, in a $\sim$4~h wide band of LT centered on $\sim$21~h, $f_r$ is essentially constant with radial distance beyond the core region,at its maximum value of $\sim$2.5, while at other LTs throughout the nightside $f_r$ is maximal in a narrow range of radial distances centered on $\sim$15~\RS\ and declines at larger radial distances.
Meanwhile, on the dayside $f_r$ remains almost constant at near-unit values.
Once again, the spatial variation evident in $f_r$ does not appear to be the result of sampling effects, as regions of enhanced $f_r$ are encountered at intervals widely spaced in time throughout the mission, appreciated again by comparing with Figure~\ref{fig:traj}.

Finally, the relative amplitudes of the azimuthal component $f_\varphi$ exhibit variations with radial distance and LT that have some similarities to those seen in both $f_r$ and $f_\theta$.
Specifically, near-unit values are found in the core region, where $\overline{f}_\varphi=0.85\pm0.30$, somewhat lower overall than the equivalent values $\overline{f}_r$ and $\overline{f}_\theta$.
However, the maximum values of $f_\varphi$ are reached at radial distances outside the core region on the nightside, as was the case for the radial component.
Furthermore, relative amplitudes $f_\varphi$ remain elevated towards the outer boundary of the region studied in a narrow range of LT centered on $\sim$21~h, though to a lesser degree than was the case for $f_r$ in the same region.

We turn now to the corresponding plots of the relative phase variation $\xi_i$ shown in both Figures~\ref{fig:str}g-i and j-l for each field component as indicated.
Panels g-i show relative phases across the equatorial plane versus radial distance in the same four quadrants of LT used above in panels a-c, whilst panels j-l show relative phases color coded using a cyclical color table.
Relative phase values $\xi_i$ close to 0$^\circ$ indicate regions where the field oscillations are in-phase with those of their respective core field component.
Values increasing away from zero in the positive direction indicate an increasing lag of the local oscillation compared to the core region.
Commenting first on the phases determined in the core region, we indeed find average phases $\overline{\xi}_i$ that are close to zero in all components.
Specifically, $\overline{\xi}_r=-3.8\pm21^\circ$, $\overline{\xi}_\theta=-9.6\pm22^\circ$ and $\overline{\xi}_\varphi=-4.6\pm17^\circ$, with these values also given in Table~\ref{tab:stuff}.
Corresponding estimates of the standard error of these quoted mean values are $\sim$$2.0-2.5^\circ$, and are given by the standard deviations reduced by a factor of $\sqrt{N}$ where $N\approx70$ is the number of populated bins within the core region ($\sqrt{70}\approx8.4$).
As can be seen from the data shown in Figures~\ref{fig:str}d-f, systematic variations of $\xi_i$ with LT inside of $\sim$12~\RS~are indeed present in all components, indicating modest departures from a purely quasi-uniform field, as was reported previously by~\citet{andrews10a}.
This is most evident in the radial component, where $\xi_r$ is generally positive at LTs towards 03~h, and generally negative at LTs towards 15~h within the core region, with the magnitude of the deflection from zero as large as $\sim$45$^\circ$.

For the radial and colatitudinal components, the relative phases $\xi_r$ and $\xi_\theta$ are seen to increase steadily to larger positive values with radial distance at most LTs, both inside and outside of the core region, indicative of radially outward propagation of the phase fronts of the oscillations.
A notable exception is however apparent in $\xi_r$ in the post-noon sector, where the phase relative to the core apparently increases to larger negative values with increasing distance.
Further exceptions to this general behaviour are often present in the outermost bins on the dayside, where significant spatial `scatter' in the phases can be seen, adjacent bins often having values discrepant with their immediate neighbours.

Meanwhile, the azimuthal component exhibits a rapid increase in relative phase $\xi_\varphi$ outside of the core region, at radial distances centered on $\sim$15~\RS.
The magnitude of this phase increase is largest at $\sim$03~h LT, and is cumulatively $\sim$180$^\circ$ over an interval of $\sim$6~\RS\  spanning $\sim$12 to $\sim$18~\RS, thus leading to a complete reversal of the sense of the azimuthal PPO field component over this range.
This phase increase is associated with a shear in the azimuthal field produced by the most significant PPO field-aligned currents, as will be shown below.
However, the magnitude of this azimuthal component phase increase is somewhat reduced at LTs from $\sim$09~h to $\sim$21~h (via dusk), and indeed is entirely absent from the measurements in the afternoon quadrant centered on 15~h~LT.

\begin{figure}[htp]
  \includegraphics[width=\textwidth]{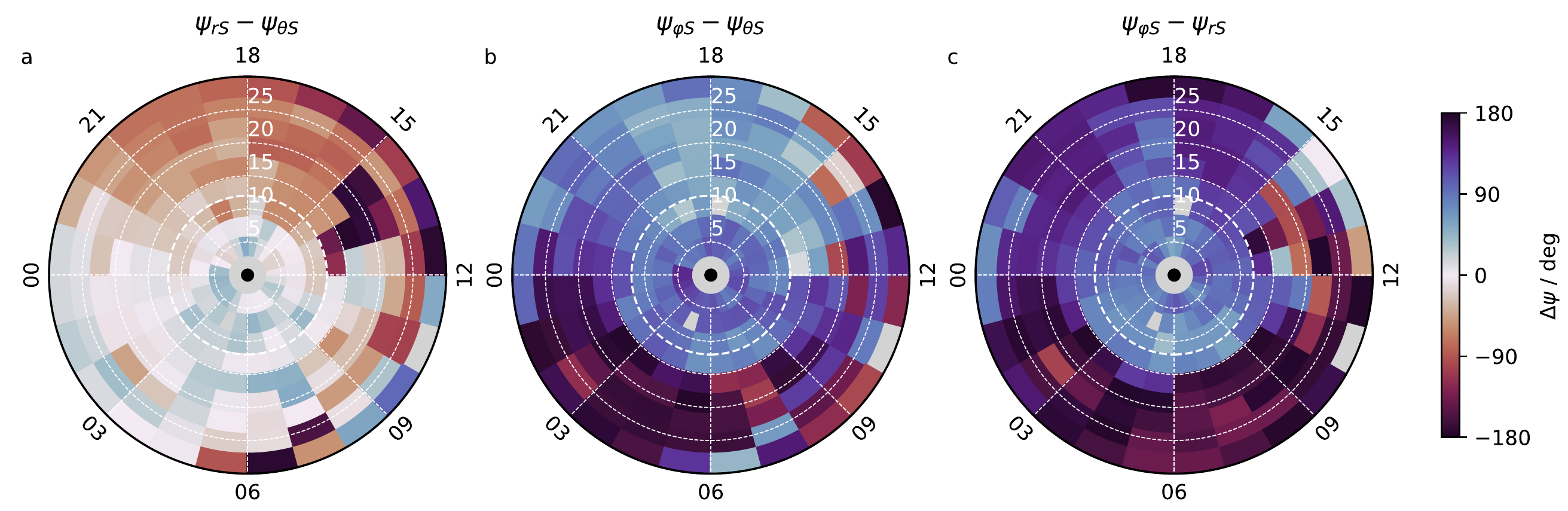}
  \caption{Plots showing the relative phases of the three field components in the equatorial plane, in a format related to the southern PPO system, and otherwise similar to that used in Figure~\ref{fig:str}.
  Panels a - c show the differences between pairs of spatial phase components $\psi_i$, as indicated in each panel, determined from fits to magnetic field data in each bin.
  Adding (or subtracting) 180$^\circ$ to the data shown in panels a and b produces the equivalent quantities northern PPO system.
  }
  \label{fig:pol}
\end{figure}

The relative phases of the three field components are shown colour-coded in Figure~\ref{fig:pol}, presented in a format appropriate to the southern hemisphere system.
An entirely equivalent set of plots appropriate to the northern hemisphere is obtained by the addition of a constant $\gamma_{\theta N} - \gamma_{\theta S} = 180^\circ$ to the values shown in Figures~\ref{fig:pol}a and b.
Figure~\ref{fig:pol}c is identical for both the northern and southern hemisphere systems.
Calculating first the component phases $\psi_{iS} = \xi_i + \gamma_{iS}$,
Figures~\ref{fig:pol}a, b, and c show the phase differences $\Delta\psi_{r-\theta S} = \psi_{rS} - \psi_{\theta S}$, $\Delta\psi_{\varphi-\theta S} = \psi_{\varphi S}- \psi_{\theta S}$, and $\Delta\psi_{\varphi-r S} = \psi_{\varphi S}- \psi_{r S}$, respectively, determined in the same spatial bins used for Figure~\ref{fig:str}.
Considering first the core region $r < 12$~\RS, the data shown in Figure~\ref{fig:pol} again clearly show results consistent with those reported previously, in which the radial and colatitudinal components are close to in-phase, $\overline{\Delta\psi}_{r\theta} = 5.8\pm 26^\circ$.
Meanwhile, the azimuthal component is in lagging quadrature with both the radial and colatitudinal components in this region, $\overline{\Delta\psi}_{\varphi\theta} = 95.0\pm 18^\circ$ and  $\overline{\Delta\psi}_{\varphi r} = 89.2\pm 16^\circ$.
This approximate lagging quadrature of the azimuthal component is maintained at larger distances in the afternoon and evening sectors of the magnetosphere.
However, the rapid change in the phase of the azimuthal field oscillation that takes place across radial distances $r \approx 15$ ~\RS\ and LTs from approximately midnight to pre-noon (through dawn) breaks this lagging quadrature, instead producing a leading quadrature relationship.
No such rapid, localized relative phase changes are observed in any of the other components.

\subsection{Radial and azimuthal phase velocities}\label{sub:vel}

The spatial variations of the relative phases $\xi_i(r,\varphi)$ discussed in the previous section necessarily imply the propagation of phase fronts, surfaces of constant oscillation phase, through the system during the oscillation cycle.
The typically increasing values of relative phase $\xi_i$ with increasing radial distance noted in section~\ref{sub:str} is indicative of outward propagation of these phase fronts, as found previously by~\citet{andrews10a}.
Meanwhile, LT variations of $\xi_i$ at a given radial distance indicate asymmetric propagation speeds of these phase fronts in the azimuthal direction.
In this section, we calculate phase velocities from the profiles of relative phase $\xi_i$ presented in the previous section.

\begin{figure}[tp]
  \includegraphics[width=\textwidth]{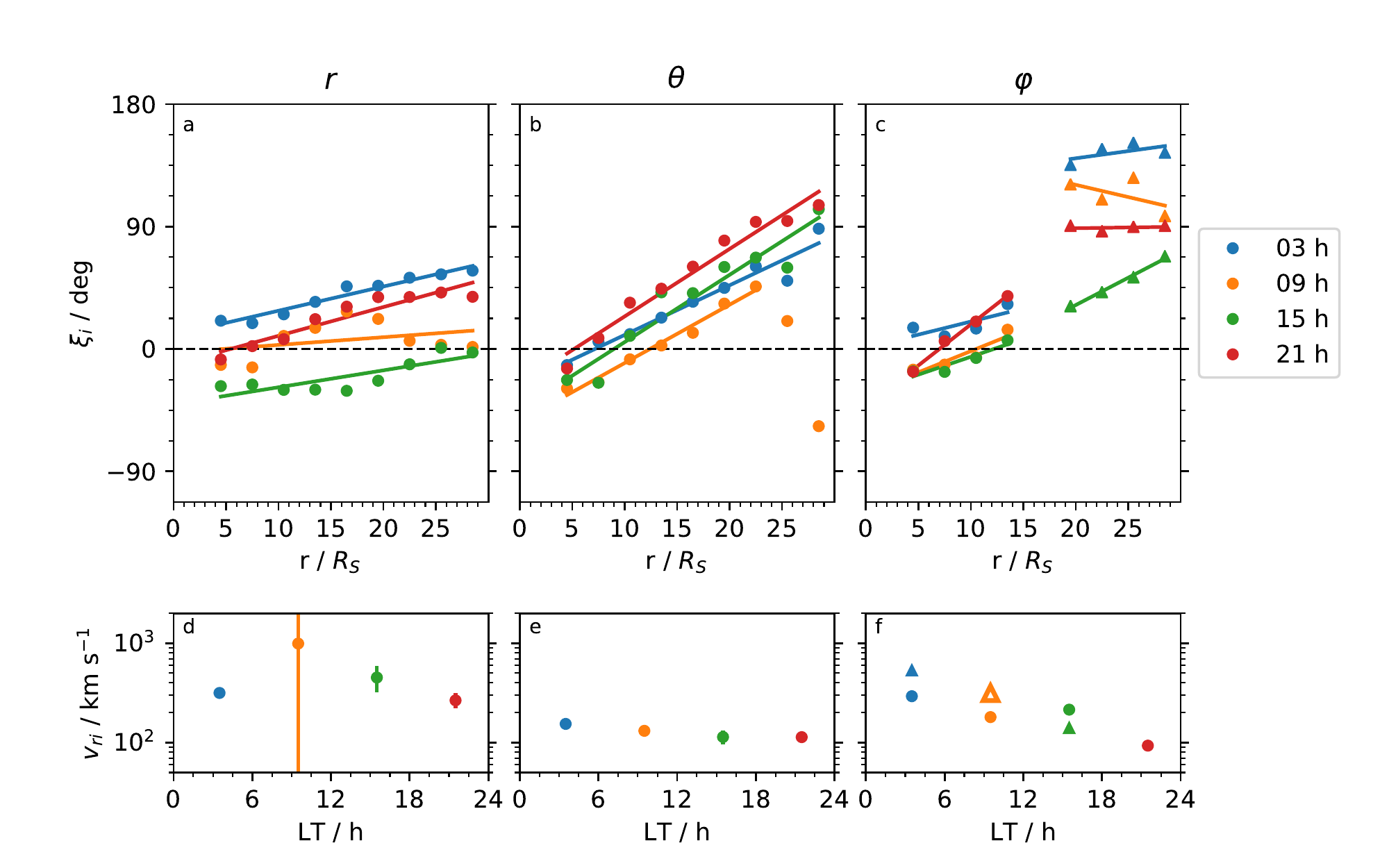}
  \caption{Panels a-c: relative phases $\xi_i$ plotted versus radial distance computed in four LT quadrants centered at 03, 09, 15 and 21~h for each component $i$, as per Figure~\ref{fig:str}g-i.
  Linear fits to data at each LT are shown by the solid lines.
  For the specific case of the azimuthal component, two fits are performed over separated radial ranges, $r < 15$~\RS\ and $r > 18$~\RS.
  Panels d-f: Radial phase velocities $v_{ri}$ and uncertainties, computed using equation~\eqref{eqn:vr} using the linear fits shown in panels a-c.  Triangular symbols in panel~f indicate those fits obtained in the outer radial range, and open symbols indicate negative values of velocity obtained.  Meaningful uncertainties cannot be obtained for $v_{r\varphi}$ given the small number of points in each fit.}
  \label{fig:vr}
\end{figure}

Taking the approach presented by~\cite{andrews10a}, and combining all phase terms into a single quantity,
$\Psi^*_{iN\!,S} = \Psi_{N\!,S}(\varphi,t) - \gamma_{iN\!,S}-\xi_i(r,\varphi)$,
the rate of change of phase in the frame of an observer at a point $(r, \varphi)$ on the equatorial plane moving with velocity $\vec{v} = v_r \hat{r} + v_\varphi \hat{\varphi}$, is given by
\begin{equation}
  \frac{d \Psi^*_{iN\!,S}}{dt} = \frac{\partial \Psi^*_{iN\!,S}}{\partial t} - \vec{v}\cdot\nabla\Psi^*_{iN\!,S} = \Omega_{N\!,S} - v_r\frac{\partial\xi_i}{\partial r} - \frac{v_\varphi}{r}\Big(1 + \frac{\partial\xi_i}{\partial\varphi}\Big),
  \label{eqn:dpsidt}
\end{equation}
where
\begin{equation}
  \Omega_{N\!,S} = \frac{d \Phi_{N\!,S}}{dt} = \frac{360^\circ}{\tau_{N\!,S}}.
  \label{eqn:omegan}
\end{equation}
An expression for the radial phase velocity $v_r$ is then obtained by setting $v_\varphi=0$ and $\frac{d \Psi^*_{iN\!,S}}{dt}=0$ and solving equation~\eqref{eqn:dpsidt}, giving the radial velocity of a point moving at constant $\Psi^*_{iN\!,S}$.
Explicitly, the radial phase velocity of component $i$ is given by
\begin{equation}
  v_{ri} = \frac{360^\circ}{\tau_{N\!,S}} \Big(\frac{\partial \xi_i}{\partial r}\Big)^{-1},
  \label{eqn:vr}
\end{equation}
and therefore depends on the instantaneous rotation period $\tau_{N\!,S}$ of the PPO fields in the appropriate hemisphere.
However, per its definition, the relative phase $\xi_i$ is taken to be common to both PPO systems, and hence gradients in this quantity do not depend on the system in question.
In Figure~\ref{fig:vr} we show radial phase gradients and resultant values of $v_{ri}$ computed at the four LT quadrants as in Figures~\ref{fig:str}g-i.
Panels a-c of Figure~\ref{fig:vr} show relative phases $\xi_i$ plotted versus radial distance for each component $i$, respectively, and least-squares linear fits to these data at each LT.
For the azimuthal component data shown in Figure~\ref{fig:vr}c, relative phases are split into two radial ranges at each LT, specifically those obtained within 15~\RS\ and those obtained outside of 18~\RS.
We discount data taken in the intermediate range, 15-18~\RS\  and thereby remove the contribution of the large rapid $\sim$180$^\circ$ phase shift occurring across this range, performing instead two separate fits to the remaining data.
The lower panels of Figure~\ref{fig:vr} show radial phase velocities $v_{ri}$, computed from the gradients of the linear fits in the upper panels using equation~\eqref{eqn:vr}, with points plotted in the appropriate color.
For the purposes of presentation, we compute $v_{ri}$ using a single constant value of the rotation period $\tau=10.7$~h rather than showing values specific to a single hemisphere at some instant of time.
Differences in radial phase velocities between those shown in Figure~\ref{fig:vr} and those calculated at any other interval for a specific hemisphere, and therefore some other specific value of $\tau_{N\!,S}$ are at most $\pm$1\%.
In Figure~\ref{fig:vr}f for the azimuthal component, two values of $v_{r\varphi}$ are therefore shown, with circles and triangles indicating values obtained over the inner and outer radial range values, respectively.
Open symbols denote negative values of $v_{ri}$, i.e.\ inward rather than outward phase propagation, and are found only in the outer radial range in the azimuthal component.
Values of the radial phase velocities, averaged over all LT, $\overline{v}_{ri,N\!,S}$ are given for each component in Table~\ref{tab:stuff}.
For the case of the azimuthal component, the average is computed only over the innermost region of radial distances shown in Figure~\ref{fig:vr}f.
To aid comparison with other studies, equivalent radial phase gradients in units of $^\circ\mathrm{R_S}^{-1}$ are given in parentheses in Table~\ref{tab:stuff}.
These values are simply the gradients of the linear fits shown in Figure~\ref{fig:vr}a-c.

Discussing first the radial phase velocities of the colatitudinal component $v_{r\theta}$ shown in Figure~\ref{fig:vr}e, derived from relative phases $\xi_\theta$ shown in Figure~\ref{fig:vr}b, typical values of $v_{r\theta}$ of $\sim$100~$\mathrm{km\;s^{-1}}$ are found, which vary only by a factor of $\sim$2 with LT.
We note that the two outermost phases $\xi_\theta$ determined in the 09~h~LT sector are not included in the linear fit, as they are significantly off the trend established at smaller radial distances.
As discussed previously in relation to Figure~\ref{fig:str},  relative phases determined in this region are likely influenced by external factors, having small amplitudes and phases that are discrepant when comparing adjacent bins.
Meanwhile, the radial phase velocities of the radial component oscillation, $v_{rr}$ are everywhere elevated compared to those of the colatitudinal component, having typically values of $\sim$300~$\mathrm{km\;s^{-1}}$.
A more pronounced variation with LT is also apparent, with relative phases almost constant with radial distance at 09~h~LT, indicative of faster outward propagation post-dawn than at other LTs.

Radial phase velocities of the azimuthal component $v_{r\varphi}$ determined in the range $r < 15$~\RS\ show little appreciable variation with LT, having values that are comparable to $v_{r\theta}$ in magnitude and positive, again indicative of outward propagation of phase fronts.
Then, at larger radial distances $r > 18$~\RS, beyond the localised rapid phase increase associated with the reversal in $B_\varphi$ in the pre-midnight to post-dawn sector, relative phases $\xi_\varphi$ apparently decrease with increasing radial distance in all except the post-dusk LT sector, indicative of inward propagation of phase fronts ($v_{r\varphi}$ negative).
However, the sub-division into two radial ranges and the corresponding reduction in the number of relative phase measurements available does not lend great confidence to these results.

\begin{figure}[htp]
  \includegraphics[width=\textwidth]{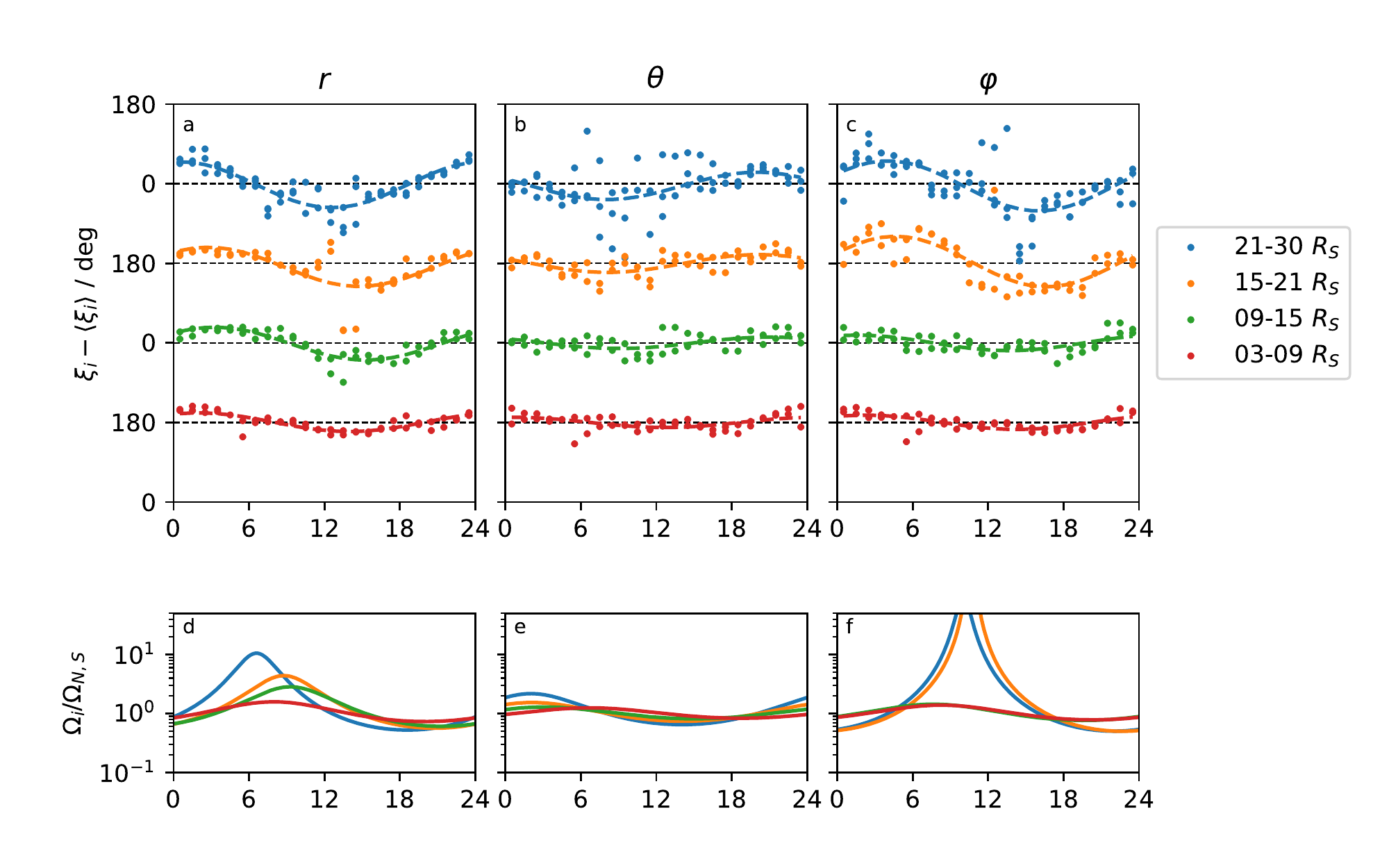}
  \caption{Panels a-c show relative phases $\xi_i$ versus LT for field component $i$, respectively, in four groups of radial distance as indicated in the legend to the right.  In each group, an arbitrary constant is added to the phase measurements to separate them over the page, and black dashed lines indicate their respective zeroes.
  Colored dashed lines meanwhile show sinusoidal fits to the data in each group.
  Panels d-f show azimuthal phase velocities determined in each component $i$ from the fits shown in the corresponding panels a-c.}
  \label{fig:vphi}
\end{figure}

Corresponding profiles of azimuthal phase velocity are obtained by instead setting $v_r=0$ and solving $\frac{d \Psi^*_{iN\!,S}}{dt}=0$ for $v_\varphi$, yielding an expression for the azimuthal phase velocity $\Omega_i$ relative to the `system' rotation rate $\Omega_{N\!,S}$,
\begin{equation}
  \frac{\Omega_i}{\Omega_{N\!,S}} = \Big(1 + \frac{\partial \xi_i}{\partial \varphi}\Big)^{-1}.
  \label{eqn:vphi}
\end{equation}
Figure~\ref{fig:vphi} shows relative phases $\xi_i$ and determined azimuthal phase velocities $\Omega_i / \Omega_{N\!,S}$.
In Figure~\ref{fig:vphi}a-c we show $\xi_i$ for each component $i$, grouped into four separate radial ranges; 3-9, 9-15, 15-21 and 21-30~\RS.
These data were initially displayed in Figure~\ref{fig:str}j-l, but here have an additional arbitrary constant phase offset added in each radial range for presentational purposes.
Sinusoidal least-squares fits are performed to each group of phase data, and are shown by the colored dashed lines.
Only variations with azimuthal wavenumber $m=1$ are considered, these being found to reasonably approximate the variations in the data that are most evident.
From these fits, equation~\eqref{eqn:vphi} is evaluated and resulting relative phase velocities plotted in Figure~\ref{fig:vphi}d-f, using the same color scheme.
Again, we proceed by discussing the results obtained for the colatitudinal component, shown in Figure~\ref{fig:vphi}b and e.
Here, relatively little variation in phase is noted with LT at radial distances less than $\sim$15~\RS\ (red and green points and lines in these panels), with the result that the relative azimuthal phase velocity $\Omega_\theta/\Omega_{N\!,S}\approx1$ and independent of LT in this range.
At larger distances, an $m=1$ variation becomes apparent despite the simultaneously increasing scatter in the relative phases (orange and blue points and lines), and a profile of relative azimuthal phase velocity that peaks in the pre-noon sector is obtained.
Phase fronts thus progress through azimuth (LT) fastest post-midnight, ahead of the nominal rotation rate of the system, whilst being retarded relative to the system rotation rate in the opposing post-noon-midnight sector.
Peak velocities are of the order of $1.4-2$ times the  system rotation rate, with corresponding reductions at opposing LTs.

Turning to the remaining two (transverse) components, with relative phases $\xi_r$ and $\xi_\varphi$ shown in Figures~\ref{fig:vphi}a and c, larger variations with LT are seen than in the colatitudinal component.
These correspond to larger departures from the system rotation rate, as computed using equation~\eqref{eqn:vphi} and shown in Figures~\ref{fig:vphi}d and f.
In each case, phase fronts are seen to exceed the system rotation rate through the pre-dawn to post-noon sector, to progressively larger degrees with increasing radial distance.
Correspondingly, in the opposing LT sector spanning pre-dusk to post-midnight the phase fronts are retarded relative to the system rotation rate.
We note that at the locations of the maxima in $\Omega_i / \Omega_{N\!,S}$, the magnitude of the departure from unity is rather sensitive to the exact parameters of the sinusoidal fit to the underlying relative phase data, and hence only the broad regions of departure from near-unit values in Figures~\ref{fig:vphi}d-f should be considered of consequence.

\subsection{Equatorial field reconstruction}\label{sub:eq}

\begin{figure}[htp]
  \includegraphics[width=\textwidth]{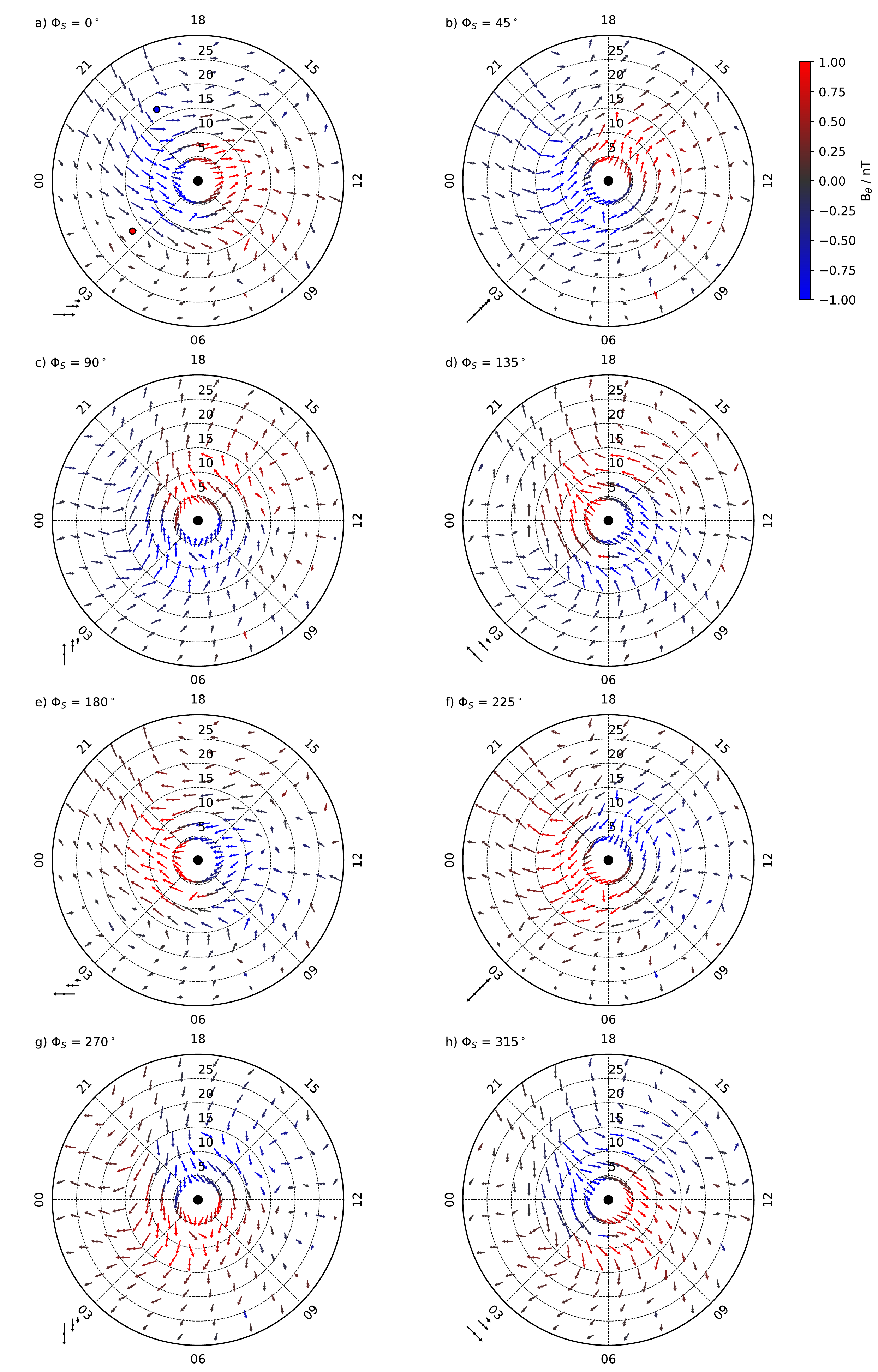}
  \caption{Caption next page.}
\end{figure}
\addtocounter{figure}{-1}
\begin{figure}
\caption{Plots showing equatorial magnetic field oscillation vectors derived from fits of equation~\ref{eqn:commonfit} to data, evaluated at phases $\Phi_S=0^\circ, 45^\circ, \ldots 315^\circ$ and presented in a format appropriate for the southern hemisphere system.
  Vectors are calculated using equation~\eqref{eqn:bis_mv}, and are thus shown specifically for the southern PPO system, using averaged core region amplitudes shown by the dotted lines in Figure~\ref{fig:ts}e-g.
  Each panel a-h depicts the transverse component ($r$-$\varphi$) within each bin by an arrow along the field direction at the phase $\Phi_S$ indicated.
  The arrows are non-linearly scaled, with the icons in the lower-left of each panel indicating 1, 2, and 5~nT vectors oriented along the expected instantaneous direction of the transverse core region field.
  The colatitudinal component is color-coded using the color table, such that positive values, shown by red colors, are into the plane of the page (and negative blue outward).
  Each panel is labelled according to the oscillation phase $\Phi_S$, and is spaced by $45^\circ$ from the previous.
  A corresponding set of plots appropriate to the northern hemisphere system can be approximated by inverting the sense of the transverse component, i.e.\ inverting the colour table used to display the $B_\theta$ component.
  Red and blue circles in panel~a indicate the approximate central locations of the two regions of curl discussed in the text.
  }
  \label{fig:eq}

\end{figure}

The fits determined above fully describe the amplitude and phase of the equatorial PPO magnetic field throughout the oscillation cycle, within each spatial bin.
We are therefore able to reconstruct the oscillating field throughout the equatorial plane, at arbitrary rotation phase $\Phi_{N\!,S}$, such that the full structure of the rotating fields can be appreciated.
This is shown in Figure~\ref{fig:eq}, in a format similar to that used by~\citet{andrews10a}, and in Figure~\ref{fig:indep_ns}.
Here, we show field vectors appropriate to the southern PPO system, plotted at eight intervals of southern phase $\Phi_S$ throughout the complete cycle.
Using the results obtained from the fits to equation (6) one could in principle compute field patterns separately for each equatorial temporal interval using the individual southern system core region amplitudes $B_{iS0}$ shown in Figures~\ref{fig:ts}e-g.
However, because the ratios of the core amplitudes for each field component are approximately constant from interval to interval, with, e.g., the radial component being approximately half the azimuthal, to a good first approximation the patterns remain fixed over time and simply scale in magnitude with the overall southern system amplitude.
Here, therefore, we instead use the overall averaged southern core amplitudes $\overline{B}_{iS0}$ shown by the red dotted lines in Figures~\ref{fig:ts}e-g, and given in Table 1.

The plotted vectors are then given by
\begin{eqnarray}
  B_{iS}(\Phi,r,\varphi) &=& f_i(r,\varphi) \overline{B}_{iS0} \cos(\Phi_S - \varphi - \xi_i(r,\varphi) - \gamma_{iS})\nonumber\\
  &=& f_i(r,\varphi) \overline{B}_{iS0} \cos(\Phi_S - \varphi - \psi_{iS}(r,\varphi)),
  \label{eqn:bis_mv}
\end{eqnarray}
where the values $f_i$ and $\xi_i$ are those determined from fits to field data within each bin, for each component $i$.
We note that since the ratios of the core amplitudes for each field component for the northern system is the same as those for the southern system for each interval (equal $k$ value), with near-equal overall averaged values as well, the field vector pattern for the northern system relative to $\Phi_N$ given by an expression corresponding to equation~\eqref{eqn:bis_mv} will be almost the same as that for the southern system shown in Figure~\ref{fig:eq}, except for a reversal in the color table showing the colatitudinal field.

Figure~\ref{fig:eq}a depicts the oscillation field at $\Phi_S=0^\circ$.
This corresponds to the point in the cycle where the `quasi-uniform' core field ($r < 12$~\RS) is expected to point towards the Sun.
Within the core region, field vectors need show the expected quasi-uniform structure, each individual plotted vector oriented approximately towards noon.
At larger radial distances ($r > 12$~\RS) the field becomes that of a transverse dipole, the dipole being oriented towards noon at this phase.
However, as was noted previously by~\citet{andrews10a}, despite the overall agreement there are nevertheless clear differences in the observed structure of the field oscillations from those depicted in the simple model, e.g.\ in Figure~\ref{fig:ppo}.
For example, a reduction in the magnitude of the field is evident in the innermost bins for which it is computed, suggestive of exclusion of this field from the innermost regions of the magnetosphere at radial distances below $\sim$3~\RS.
There is also a clear tendency of the innermost ring of field vectors to be `deflected' around the planet as a consequence of the low values of $f_r$ (and therefore $B_r$) in this region.
At larger distances, towards the outer edge of the core region in the midnight sector ($\sim$21-03~h LT), the transverse field vectors can be seen to be displaced relative to the nominal core region quasi-uniform field, and `lag' in the sense of rotation by $\sim45^\circ$.
Beyond the core region, the rapid shear in the phase of the azimuthal component that takes place at distances beyond $\sim$15~\RS\ in the $\sim$21 to $\sim$09~h (via midnight) LT range is here seen to correspond to two distinct and extended regions of circulation (curl) in the field, one centered in the post-dusk sector and the other most evident in the dawn sector.
To guide the eye, we indicate the approximate center of each region of curl by the blue and red filled circles in Figure~\ref{fig:eq}a.
The two regions of field curl, having opposite senses (clockwise at dawn, counter-clockwise at dusk) correspond to two distinct regions of electrical current crossing the equatorial plane, as discussed below.
However, these two regions of curl are of markedly different form when compared at dawn and dusk at the instant shown, with the reversal of the transverse field in the dawn sector occurring over a narrower range of radial distance and a more extended range of LT than its dusk sector counterpart.
This necessarily implies that the strength of the currents flowing through the equatorial plane is also asymmetric.
Such a pronounced dawn-dusk asymmetry, as found originally by~\citet{andrews10a} but nevertheless also present in this much larger survey, is not explained by the simple LT-independent theoretical description of PPO fields and currents given in section~\ref{sec:field}.

Following the evolution of the field through the oscillation cycle, through $\Phi_S=45^\circ$ to $\Phi_S=90^\circ$ (Figures~\ref{fig:eq}b and c), the steady rotation of the core region field in the sense of planetary rotation  is evident, along with the motions of the two distinct regions of field circulation (counter-clockwise in this Figure).
The region of field circulation originally present in the post-dusk sector, having a counter-clockwise sense, grows in its intensity through this part of the cycle, the magnitude of the deflection in the field through this region increasing as it moves into the pre-midnight sector.
At the same time, the region of opposite sense field circulation found in the dawn-sector weakens, becoming distributed over a larger radial extent as it rotates into and through the dayside magnetosphere, more rapidly than its dusk-side counterpart.

At the following instant depicted, $\Phi_S=135^\circ$ (Figure~\ref{fig:eq}d), the weaker, more extended clockwise circulation of the field moves from the dayside into the dusk sector and now grows in intensity through this part of the cycle.
At the same time, the counter-clockwise circulation region moves into the pre-dawn sector and its radial extent is reduced, suggesting intensification of the current threading this region.
The field continues to evolve, and at $\Phi_S=180^\circ$ the configuration is (necessarily) identical and opposite to that at $\Phi_S=0^\circ$, with the sense of each field component reversed.
The relative motions of the different structures evident in the equatorial field then evolve from $\Phi_S=180^\circ$ to $\Phi_S=360^\circ=0^\circ$ in precisely the same manner as already described from $\Phi_S=0^\circ$ to $\Phi_S=180^\circ$, albeit with the senses of all components of the field reversed (along with the sense of any circulation present).
In Movie S1, included as supporting material, we show the evolution of the equatorial field throughout the whole cycle at smaller phase increments.

Comparing these results with those obtained previously by~\citet{andrews10a}, we find that the large-scale structure of the rotating fields is clearly comparable throughout the oscillation cycle.
This is expected given the commonality of the northern and southern system oscillations found in section~\ref{sub:indep}, and the high degree to which the latter was dominant over the former during the interval analysed by~\citet{andrews10a}.
Nevertheless, the improved coverage afforded by including data obtained in the later stages of the mission is clear, with suitable fits now obtained in $\sim$~98\% of bins outside of 3~\RS\, increased from $\sim$80\% in~\citet{andrews10a}.
Further direct comparison with their original results also highlights the much improved coverage of the dusk sector presented here, our results confirming a clear dawn-dusk asymmetry in the form of the perturbation field, not described by the simple theoretical pictures.
Similarly, the larger typical volume of field data obtained within each bin, along with improved understanding of the nature of the PPO fields and their rotation rates yields fits that are more contiguous in space, e.g.\ with better-organised variations from one bin to those immediately adjacent.

As with previous plots derived from these `common' relative amplitude and phase determinations, an equivalent northern system plot can be obtained from that shown in Figure~\ref{fig:eq} by adjusting the phase of the colatitudinal component by 180$^\circ$, or equivalently inverting the color table used to indicate the $B_\theta$ component.
Strictly, the averaged core region amplitudes $\overline{B}_{iS0}$ used must also be substituted by the equivalent northern system quantities given in Table~\ref{tab:stuff}, although this correction represents only a few \% change in the amplitudes of the individual components of the field.

\section{Inferred north-south currents}\label{sec:cur}

Fully determining the transverse ($B_r-B_\varphi$) equatorial PPO field permits the calculation of the associated north-south (colatitudinal) currents $j_\theta$, positive southward, throughout the equatorial plane.
Evaluating Amp\`{e}res law in the equatorial plane yields
\begin{equation}
  j_\theta = \frac{1}{\mu_0 r} \left( \frac{\partial B_r}{\partial \varphi} - \frac{\partial}{\partial r}(r B_\varphi)\right),
  \label{eqn:amp}
\end{equation}
where $\mu_0 = 4\pi\times10^{-7}\mathrm{H m^{-1}}$.
At radial distances less than $\sim$20\RS, where Saturn's planetary field dominates over external contributions, these currents flow approximately along the direction of the magnetic field and therefore can be considered approximately field-aligned.
At larger distances, this is not necessarily the case.

The derivatives on the right hand side of equation~\eqref{eqn:amp} can be evaluated numerically.
In this paper, principally for the sake of simplicity, we make a minor adjustment to the approach used previously be~\citet{andrews10a}, here taking instead so-called `centered-differences'.
This avoids the use of the more complex binning scheme employed by~\citet{andrews10a}, in which staggered bins were used for determining the radial and azimuthal fields prior to the calculation of currents.
Specifically, we approximate e.g.\ the azimuthal gradient in $B_r$ at point $(r, \varphi)$ as
\begin{equation*}
  \frac{\partial }{\partial \varphi }B_r(r,\varphi) \approx \frac{B_r(r,\varphi+\delta\varphi) - B_r(r,\varphi-\delta\varphi)}{2~\delta\varphi},
\end{equation*}
where $\delta\varphi$ is the azimuthal spacing in the bins, equivalent to 1~h of LT (i.e., accounting for the overlap used in this direction).
This approach also has the effect of determining the currents over a slightly larger spatial scale, and moreover appears to give more stable results, particularly with regard to the determination of (typically smaller) radial gradients in the azimuthal field.

Colatitudinal current densities computed in this manner are shown color-coded in Figure~\ref{fig:cur}, with positive values of $j_\theta$ shown red, and negative blue.
Positive (southward) or negative (northward) values therefore indicate currents directed into or out of the plane of the figure, respectively.
The format of the figure is otherwise comparable to that in Figure~\ref{fig:eq}, depicting eight different instants of rotation phase $\Phi_S$.
As the determination of current density in a given bin depends on field values in adjacent bins, the regions of the figure in which no current density could be computed (shown grey) are larger than those in which no fits to the underlying field data were obtained.
For the inner and outer-most bins (3-6\RS\ and 27-30\RS), we compute the current assuming that $\partial B_\varphi/\partial r$ tends to zero at the inner and outer edges of the region considered.
Coloured circles indicate the current-weighted mean locations of all individual upward and downward currents at each phase plotted (i.e., the ``center of mass'' of each current system).

\begin{figure}[htp]
  \includegraphics[width=\textwidth]{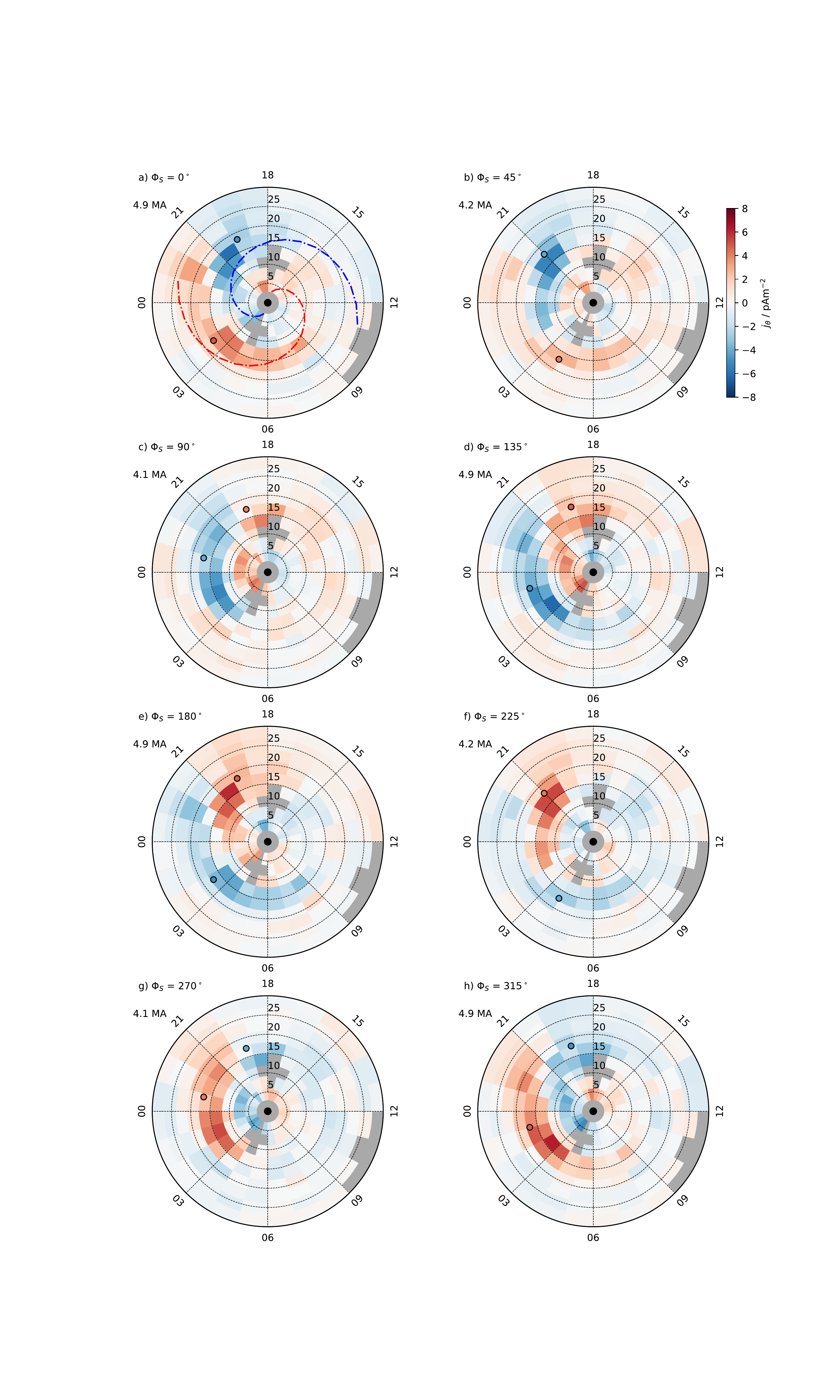}
  \caption{Caption next page.}
\end{figure}
\addtocounter{figure}{-1}
\begin{figure}
  \caption{Plots showing the north-south current density $j_\theta$ in Saturn's equatorial plane, computed from centered differences of the transverse magnetic fields shown in Figure~\ref{fig:eq}.
  Each panel a-h shows the current density computed at an instant of (southern) oscillation phase $\Phi_S$, presented in a format which is otherwise equivalent to that used in Figure~\ref{fig:eq}.
  The current density is shown colour coded in each spatial bin and is derived from the gradient in the fields across it.
  Positive (red) values indicate currents into the plane of the figure, in the positive $\theta$ direction.
  Current-weighted average locations of the positive and negative current regions are marked by the appropriately colored small circles, while two dashed lines are shown that loosely approximate the spiral structure referenced in the text in panel a only.
  The total integrated current of either sign is given below the panel label.
  As presented, the computed current densities are equivalent (to a scale factor close to unity) in the northern and southern systems, as their averaged core amplitudes shown in Figure~\ref{fig:ts}e-g are almost the same, and their calculation does not depend on the colatititudinal component $B_\theta$.}
  \label{fig:cur}
\end{figure}

Current densities shown in Figure~\ref{fig:cur} are calculated specifically for the southern hemisphere PPO system.
However, as equation~\eqref{eqn:amp} does not depend on the colatitudinal field $B_\theta$, the calculated current density profile $j_\theta$ is approximately the same for both the northern and southern PPO systems within this analysis.
The only difference arises due to the marginally different ratios of the time-averaged amplitudes $\overline{B}_{iS0}$ and $\overline{B}_{iN0}$ used, specifically the slight departures from unity of the ratios  $\overline{B}_{rS0} / \overline{B}_{\varphi S0}$ and $\overline{B}_{rN0} / \overline{B}_{\varphi N0}$.
These ratios are equivalent at the $\sim$5\% level, well below the uncertainties in their respective values and the uncertainties in the fitted field amplitude ratios in individual bins.
Consequently, the current density profile shown in Figure~\ref{fig:cur} can also be taken to represent the northern system at the rotation phase indicated $\Phi_N=\Phi_S$ to a good approximation.

We first discuss first the currents computed at the start of the cycle $\Phi_S=0^\circ$ shown in Figure~\ref{fig:cur}a.
Two regions of current, one of positive and one of negative sense, can be clearly discerned, each associated with the regions of net circulation in the transverse magnetic field noted in the corresponding Figure~\ref{fig:eq}a.
Each region can be seen to be comprised of contiguous bins of the same sign of current, at least for those bins in which the current carried is significant.
However, this contiguity is less evident beyond $\sim$20~\RS\ on the dayside, where the current densities determined are themselves low, likely once again the result of poor fits due to external variations present in the underlying magnetic field data.
The region of negative (northward) current peaks in intensity at around $\sim$20~h LT at $\sim$17~\RS, whilst extending at lower intensities outward at earlier LTs, and inward at later LTs, forming a distinct spiral pattern.
Similarly, the region of positive (red) current, which has its peak intensities at $\sim$02~h~LT again at $\sim$17~\RS\ at $\Phi_S=0^\circ$ also forms a clear spiral pattern as it weakens away from this maximum.
Red and blue dashed lines in Figure~\ref{fig:cur}a are used as visual guides as to the approximate structure of these two spirals.
These spirals significantly overlap, however, such that the sense of the current generally reverses in sense at a radial distance of $\sim$12~\RS over a large interval of LT.
Values of the separately integrated total positive and negative currents are given in each panel of the figure.
Specifically, at $\Phi_S=0^\circ$, the total integrated current of either sign is 4.9~MA.
Boundary value effects and the presence of bins in which the current density $j_\theta$ cannot be computed result in minor differences $<0.1$~MA between the total integrated positive and negative currents at a given phase.

As the cycle progresses through $\Phi_S=45^\circ$ to $\Phi_S=90^\circ$, both current regions identified above rotate around in the equatorial plane in the sense of planetary rotation.
At the same time, the total current carried drops to $\sim$4.1~MA.
The region of negative current rotates relatively slowly through the evening sector, its averaged location only shifting by $\sim$3~h of LT through this quarter-cycle.
Meanwhile, the positive current region rotates much faster in LT, traversing the entire dayside and reaching the afternoon sector by $\Phi_S=90^\circ$, having traversed $\sim$18~h of LT.
The spiral structure of these two current systems nevertheless remains evident, despite their progression in LT at markedly different rates.
Total integrated currents remain $\sim$20\% lower during this part of the cycle than at other phases.
By $\Phi_S=135^\circ$, the positive current region is now more clearly identifiable as a contiguous region of current, centered now at midnight in the core region, and occupying the entire dusk sector at larger radial distances.
The region of peak negative current remains centered on $\sim$17~\RS\ and continues to rotate further into the dawn sector.
Total integrated currents for both regions have risen to $\sim$4.9~MA.
At $\Phi_S=180^\circ$ the spatial distribution of these currents is now (by definition) equivalent to that at $\Phi_S=0^\circ$, with the sense of current density reversed in each bin.
The evolution of each region through the remaining half-cycle is thus as described above, but with the senses of the two regions reversed.

In Movie S2, included as supporting material, we show the evolution of currents flowing through the equatorial plane as computed at smaller phase increments, to be compared with Movie S1.

Once again, in comparing with those related results shown by~\citet{andrews10a}, the spatial variation of the currents obtained in this analysis shows many familiar elements, namely the dominance of those regions of current flowing in the pre-midnight to post-dawn sectors at radial distances of $\sim$12-18~\RS.
\citet{andrews10a} decomposed their observations into six individual current regions; two associated with the innermost bins spanning 3-6~\RS\ (one of each sign), and four in the outer region, spanning the principal current regions at distances greater than $\sim$12~\RS\ (two of each sign).
Instead, we here suggest instead that one need only consider two individual current carrying regions at any phase $\Phi_S$, the improved coverage afforded in this study clearly indicating that this is the case throughout the cycle.
The original computation of the total maximal current carried through the equatorial plane made by~\citet{andrews10a} of $\sim$5-6~MA is comparable to, if slightly larger than that reported here, noting that the contribution of edge-effects and data gaps is significantly reduced in the analysis presented here.

\begin{figure}[htp]
  \includegraphics[width=\textwidth]{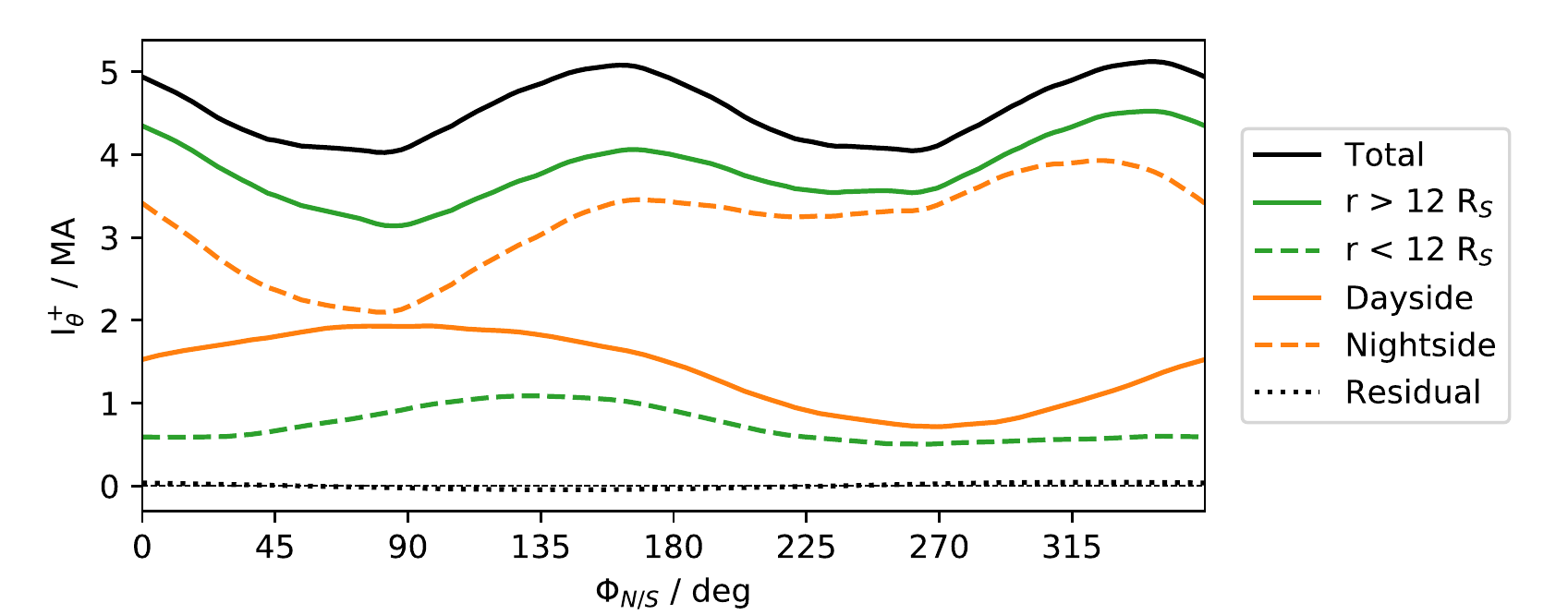}
  \caption{Integrated positive currents $I_\theta$ computed from the data shown in Figure~\ref{fig:cur}, and at intermediate phases $\Phi_{N\!,S}$.
  The solid black trace shows the total current carried of positive sign at all LTs and radial distances for a given phase $\Phi_{N\!,S}$.  The red solid and dashed lines shows the positive currents flowing at radial distances $r$ greater and less than 12~\RS, respectively.
  Blue solid and dashed lines show total positive currents flowing on the dayside and on the nightside, respectively.
  The black dotted line shows the total integrated current of both signs, i.e. the residual current flowing at given phase.}
  \label{fig:cur_ph}
\end{figure}

In Figure~\ref{fig:cur_ph} we show integrated currents in MA, computed as in Figure~\ref{fig:cur} but now at 3$^\circ$ increments of phase $\Phi_{N\!,S}$.
The `residual' integrated current, shown by the dotted black line and computed  by integrating all individual current regions irrespective of sign, can be taken to give some measure of the uncertainty on these results, as it takes non-zero values at least in part due to the treatment of the boundaries of the region studied in deriving currents, and the presence of data gaps.
Peak values of this residual or `missing' current are $\sim$0.05~MA, i.e.\ well below both the average values and the amplitude of the variations in the individual regions of positive (or negative) sign discussed below.
Hence the total positive current is equal and opposite to the total negative current at a given phase to a very good approximation.
The total positive current is shown by the solid black trace in Figure~\ref{fig:cur_ph}, and varies significantly with phase between 4.0 and 5.1~MA, with an average value of 4.5~MA across the whole cycle.
Two distinct maxima are evident in the trace, at $\Phi_{N\!,S}\approx$165$^\circ$ and $\sim$345$^\circ$, of approximately equal value.
The two interleaving minima are however somewhat different in value.
A relatively constant fraction, $\sim$80\%, of this total positive current is carried outside the core region, as shown by the solid red trace, computed only for those bins centered on radial distances greater than 12~\RS.
Those bins inside this boundary meanwhile contribute 0.5 to 1.1~MA, shown by the dashed red line in Figure~\ref{fig:cur_ph}, and average 0.7~MA over the cycle.

We note that in the most basic model of PPO current systems proposed variously by~\citet{southwood07a},~\citet{andrews10a} and~\citet{hunt14a}, the quantities shown in Figure~\ref{fig:cur_ph} are expected to remain approximately constant with rotation phase $\Phi_{N\!,S}$.
However, the departure from constant value is significant in each case, with the outermost current displaying two distinct maxima per rotation, while only a single maximum is present in the inner region currents.

In addition, we split the data into total currents carried on the dayside and nightside, shown by the solid and dashed blue lines in Figure~\ref{fig:cur_ph}, respectively.
As readily deduced from the data shown in Figure~\ref{fig:cur}, typical current densities are significantly elevated on the nightside, carrying an average of 3.1~MA over the whole cycle compared to 1.4~MA on the dayside.
Implications of this asymmetry are discussed in section~\ref{sec:discussion}.

\section{Combined northern and southern system PPO fields}\label{sec:combined}

\begin{figure}[htp]
    \includegraphics[width=0.9\textwidth]{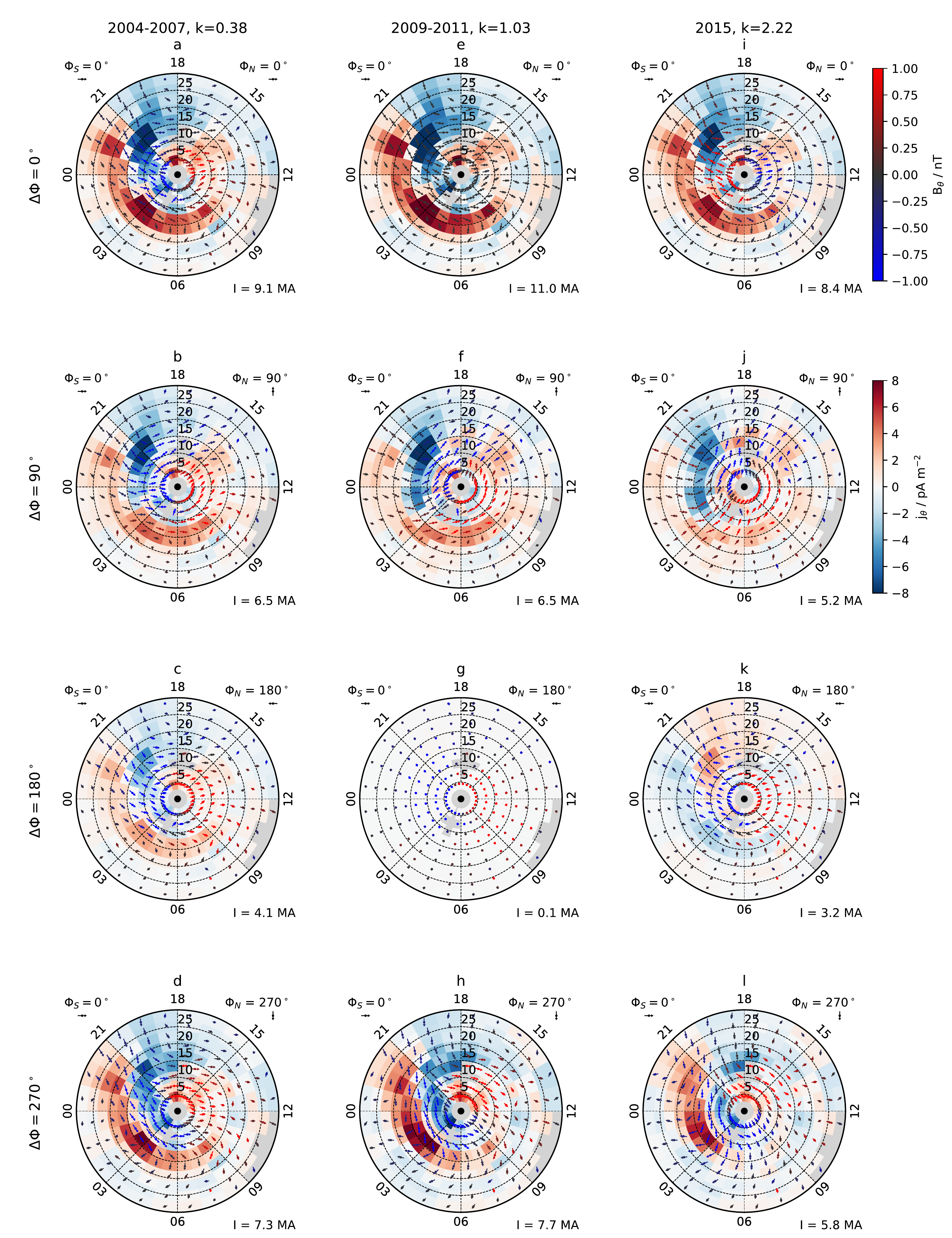}
\caption{Plots showing the effect of vectorially combining the northern and southern PPO fields and currents in the equatorial plane, for three specific epochs as shown at the top of the figure, i.e., the southern summer interval (2004-2007) in panels a-d, the post-equinox interval (2009-2011) in panels e-h, and the northern spring interval (2015) in panels i-l.
  In each case, we use the core region amplitudes for each field component as shown in Figure~\ref{fig:ts}e-g, with $k$ values as also shown at the top of this figure.
  The format of each panel is comparable to those used in Figures~\ref{fig:eq} and~\ref{fig:cur}, except that the field plotted in each is the sum of both the northern and southern oscillation, determined at the instantaneous values of $\Phi_{N}$ and $\Phi_{S}$ indicated in each panel.
  Arrows below each marked value of $\Phi_{N\!,S}$ show the expected core-field orientation of each system at this particular phase.
  For simplicity we have taken the southern phase $\Phi_S=0^\circ$ throughout, while the northern phase $\Phi_N$ is fixed at $0^\circ$ in the top row, $0^\circ$ in the second, $180^\circ$in the third, and $270^\circ$ in the bottom row.  The corresponding beat phase  $\Delta\Phi = \Phi_N - \Phi_S$ is shown to the left of each row.  }
  \label{fig:dual}
\end{figure}

As has been discussed, at any instant in time the equatorial PPO field is comprised of the sum of fields produced by the northern and southern PPO current systems.
In the preceding section, the common structure of these oscillations was determined by fitting equation~\eqref{eqn:commonfit} to field data, and presented in a format appropriate for the southern PPO system in Figures~\ref{fig:str}-\ref{fig:cur}.
Here, we briefly elucidate some features of the combined oscillations, and their variation both with the relative phase $\Delta\Phi=\Phi_N - \Phi_S$, and with the amplitudes of the core region fields, $B_{iN\!,S0}$.
To illustrate the variation in PPO fields and currents throughout the mission, in Figure~\ref{fig:dual} we show the combined northern and southern PPO fields, computed using equation~\eqref{eqn:commonfit} and the determined values $f_i$ and $\xi_i$, doing so using three different sets of core region amplitudes.
Specifically, we substitute those amplitudes determined a) during the extended interval of southern hemisphere dominance spanning 2004-2007, b) during the  interval of near-equal southern and northern amplitudes following equinox, spanning late 2009 to the beginning of 2011, and c) during the interval of northern dominance in 2015 (c.f., Figure~\ref{fig:ts}d-g).
Within each sub-panel, vectorially combined PPO field vectors are plotted using arrows in the same format used in Figure~\ref{fig:eq}.
We additionally underplot the computed field-aligned current densities obtained at the same instant using the same central-difference scheme described in section~\ref{sec:cur}, and using the same format for these data as was used previously in Figure~\ref{fig:cur}.
Each sub-panel is labelled with the values of $\Phi_N$ and $\Phi_S$ at which it is calculated.
For simplicity, $\Phi_S$ is held fixed at $0^\circ$, while each panel within a row is computed at constant $\Phi_N$, taking equally spaced values $\Phi_N=0$, 90, 180 and 270 degrees from top to bottom.
The beat phase $\Delta\Phi$ is therefore constant within each row, at the same value as $\Phi_N$, as indicated on the left of the figure.
Consequently, while the southern system oscillation is fixed in each panel, with its quasi-uniform core field pointing towards the Sun, the northern oscillation is computed at representative phase offsets from this, as indicated by the arrow icon below each labelled value of $\Phi_{N\!,S}$.
The total integrated positive current flowing through the equatorial plane, $I$, is given to the lower right of each panel.

We discuss first Figures~\ref{fig:dual}a-d, calculated using northern and southern system amplitudes $B_{iN0} = (0.33, 0.63, 0.65)$~nT and $B_{iS0}=(0.86,  1.65,  1.71)$~nT (from Figures~\ref{fig:ts}e-g), i.e.\ where the southern system is dominant over the northern with $k\approx0.38$.
Southern system amplitudes $B_{iS0}$ in this interval are thus $\sim$30\% larger than the averaged values $\overline{B}_{iS0}$ used in constructing previous figures.
Compared with the averaged values $\overline{B}_{iN0}$, the northern system amplitudes $B_{iN0}$ are reduced over this interval by a factor of $\sim$50\%.
Here, the simultaneous presence of weaker oscillations associated with the northern system is seen to have only a minor effect on the large-scale structure of the equatorial field oscillations and inferred north-south currents.
In Figure~\ref{fig:dual}a at $\Phi_N=\Phi_S=0^\circ$ the transverse fields of both systems are everywhere aligned, oscillating in phase in both systems and therefore constructively interfering and resulting in larger total amplitudes than for the southern system in isolation.
The antiphase relationship between the colatitudinal field components in the northern and southern system leads to the converse situation, with amplitudes reduced relative to the southern system value.
The enhanced transverse fields lead to an approximately proportional increase in the total (positive) current carried by factor \mbox{$\sim(1+k)B_{S0}/\overline{B}_{S0}\approx1.8$} to $\sim$9~MA, relative to the `pure' southern system value of $\sim$4.9~MA at $\Phi_S=0^\circ$.
This increase is due to the presence of the co-oriented albeit weaker northern system currents at $\Phi_N=0^\circ$ everywhere reinforcing the southern system currents.

As the northern system oscillation phase $\Phi_N$ is increased from 0$^\circ$ in (Figure~\ref{fig:dual}a) to 270$^\circ$ (Figure~\ref{fig:dual}d), the total core field orientation, and the orientation of the individual field vectors are rotated by a maximum of $\arctan(k)\approx$20$^\circ$ from their initial orientation, with this maximum deflection being reached at a beat phase of $\Delta\Phi = 90^\circ$ (and 270$^\circ$), where the two systems are in quadrature.
The amplitude of the field drops to a minimum as $\Delta\Phi$ reaches 180$^\circ$.
The central locations of the field circulations, and the peaks in the north-south current densities at larger radial distances, are similarly rotated by $\sim$20$^\circ$.
However, we note that it is precisely these small systematic shifts in field orientation about that expected for a `pure' southern system signal that allowed the detection of the presence of this northern system and the inference of its properties.
At $\Phi_N=180^\circ$ (Figure~\ref{fig:dual}c), the weaker northern PPO fields now oppose those of the dominant southern system in the transverse components, whilst reinforcing them in the colatitudinal component.
Similarly, the total current drops as $\Phi_N$ increases, reaching a minimum of $\sim$4.1~MA at $\Phi_N=180^\circ$, when the northern currents are everywhere opposite in sign to those of the southern.

Figures~\ref{fig:dual}e-h show total fields and currents computed during a situation of near-equal northern and southern system amplitudes, i.e.\ where $k$ is close to unity.
Such intervals have occurred at least twice during the Cassini epoch, namely from August 2009 to February 2011 and again from October 2012 to July 2013, the former interval being represented here.
Specifically, fields are evaluated using core amplitudes $B_{iN0} = (0.72, 1.15, 1.46)$~nT and $B_{iS0}=(0.70,  1.12,  1.42)$~nT, $k\approx1.03$.
During these two intervals for which $k$ has been explicitly measured as being close to unity, the rotation periods of the two systems were nevertheless distinct, separated by several minutes with the northern being the shorter of the two.
Consequently, $\Delta\Phi$ was not fixed during these intervals, but slowly increased with time by a few degrees per day.

Modulations in the amplitude and phase of the combined field oscillations are maximal when $k=1$, and consequently the total current flowing through the equatorial plane may also reach extremal values.
When the two systems reinforce each another at $\Delta\Phi=0^\circ$ (Figure~\ref{fig:dual}e), the transverse components of the field constructively interfere and are similarly doubled in magnitude without altering their direction, and the magnitude of the currents flowing through the equatorial plane are similarly increased.
Meanwhile, the colatitudinal fields, having opposite sense in the two systems, destructively interfere and are reduced to near-zero values.
These effects can be discerned in Figure~\ref{fig:dual}e, in which the colatitudinal component of the field is reduced to near zero (displayed vectors all black), the transverse component is increased (vectors longer than in Figure~\ref{fig:dual}a, although somewhat obscured by the non-linear scale used).
The total current current through the system is increased to $\sim$13~MA.

In Figure~\ref{fig:dual}f for $\Delta\Phi = 90^\circ$ the transverse components of the fields can be seen to be everywhere rotated towards later LTs, and lie along the mean direction of the northern and southern systems.
The core field, for example, inside of $\sim$15~\RS\ is oriented towards $\sim$15:00~h~LT ($\varphi=\Delta\Phi/2=45^\circ$) in Figure~\ref{fig:dual}f.
Meanwhile, the colatitudinal components no longer perfectly cancel between the northern and southern systems, and the total current flowing decreases towards typical values for the individual systems themselves.
When $\Delta\Phi$ reaches 180$^\circ$ as shown in Figure~\ref{fig:dual}g, and the core fields of the two systems are oppositely directed, the other extreme is reached.
In this situation, the transverse fields have almost completely cancelled one another (individual field vectors so small as to be invisible in the scale used), and as a result almost no current flows through the equatorial plane in this situation.
However, although the presentation does not emphasise this, the colatitudinal component of the field is still present, and at relatively enhanced amplitudes, as the two systems now constructively interfere with one another at this relative phase in this field component.

Finally, Figures~\ref{fig:dual}i-l show fields and currents representative of the interval of strong northern dominance in 2015, where
$B_{iN0} = (0.85, 1.66, 1.39)$~nT and $B_{iS0}=(0.38,  0.75,  0.63)$~nT, and $k\approx2.22$.
Here, we keep $\Phi_S=0^\circ$, as in the other panels in Figures~\ref{fig:dual}, such that the total fields and currents rotate about the planet with the phase $\Phi_N$ of the now-dominant northern system.

Here the presence of the weaker southern system subtly perturbs the field oscillations of the dominant northern system as it rotates, leading to small changes at the level of $\sim$25$^\circ$ in the orientation of individual vectors in the equatorial plane, and variations in the total current carried of similar magnitudes to those discussed in relation to the opposing situation depicted in the left column of Figure~\ref{fig:dual}.

\section{Discussion}\label{sec:discussion}

We now discuss the implications of the results presented here, and their connection to related observations, beginning by considering their relationship to observations of PPO field-aligned currents determined from high-latitude observations by Cassini.
Directly from Amp\`{e}re's law, the electrical currents determined in this study at a given location on the equatorial plane reflect only those PPO-related variations in the component of the current flowing north-south through the equatorial plane.
Such currents may arise as the result of two distinct current systems.
Firstly, field-aligned currents flowing from one hemisphere to the other  will in principal be accurately determined, both in location and intensity.
Secondly, currents associated with PPO related motions of Saturn's quasi-static field-perpendicular current systems will also contribute, if the modulation results in periodic deflections towards the north and south across the equatorial plane.
Specifically, both the ring current, flowing azimuthally in the sense of planetary rotation at radial distances of $\sim$7-20~\RS, and the cross-tail current flowing in the same sense beyond this at larger distances on the nightside are expected to be modulated in response to the PPOs~\citep[see e.g.][]{arridge11a, carbary13c, cowley17a, cowley17b}.

Comparing results presented here to studies of field-aligned currents modulated by PPO phase, we note that the most intense currents found in this study at $\sim$15~\RS\ in the equatorial plane lie on field lines approximately co-located with Saturn's main auroral oval, at latitudes of $\sim$70$^\circ$~\citep{bunce08a}.
Perturbations in the high latitude magnetic field due to these currents, principally in the azimuthal component, have been studied independently of the equatorial data, PPO modulations in these observations first having been noted by \citet{talboys09a} and \citet{talboys09b} with the first detailed analysis by~\citet{hunt14a}, and recently extended by~\citet{bradley18a}.
Analyses of data obtained by Cassini on high-latitude orbits has suggested currents of the order of $\sim$2-3~MA per half-cycle during the 2008 and 2012-2013 intervals~\citep{hunt14a, bradley18a}, and further inferred at least partial inter-hemispheric closure of these currents~\citep{hunt15a}.
These estimates of the total current should be compared to those given in this study for the average total current flowing through the equatorial plane at radial distance $r \ge 12$~\RS\ of $\sim$4~MA, i.e.\ showing a 25-50\% discrepancy.
Those departures from the simple theoretical description of PPO currents noted in the results presented in this paper, for example in the variation in the total current carried with oscillation phase (Figure~\ref{fig:cur_ph}) are not sufficient to explain this discrepancy.
However, considering only the currents flowing through the equatorial plane on the dayside, the peak positive current is $\sim$2~MA (occurring at $\Phi_{N\!S}\approx0^\circ$ to first order), much closer to the 2-3~MA inferred from high-latitude observations.
The presence of some amount of cross-field closure of the high-latitude currents, such as that proposed e.g.\ by~\citet{hunt14a}, would further improve the correspondence.
The large discrepancy between the total PPO-related currents flowing on the dayside as compared to the nightside is indicative of a significant additional contribution due to the periodic motion of the cross-tail currents occurring tailward of the dawn-dusk meridian.

\begin{figure}
  \includegraphics[width=\textwidth]{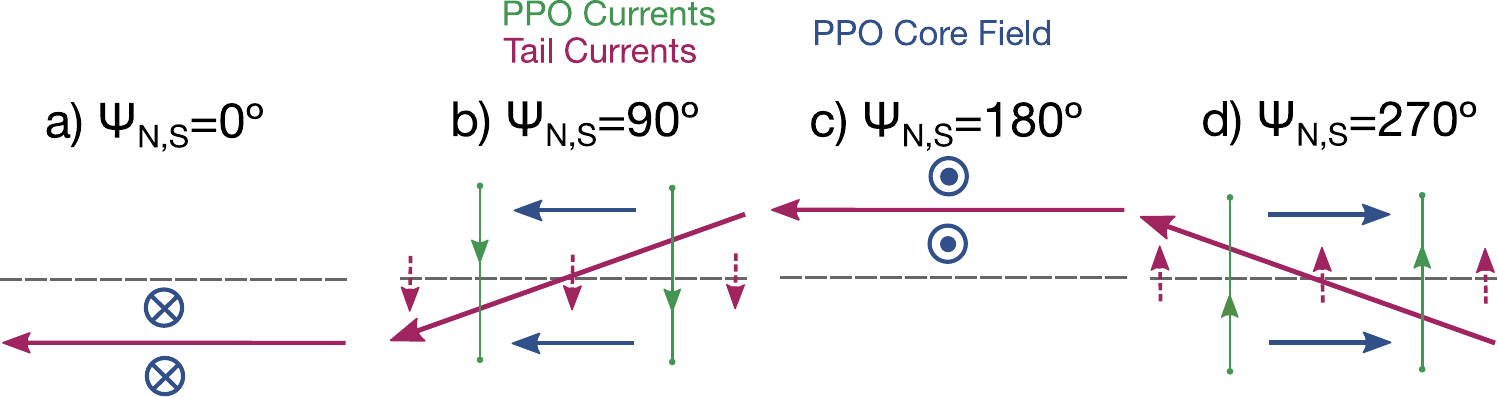}
  \caption{Schematic illustration of the north-south current resulting from the PPO-related periodic motion of the plasma sheet and cross tail current, as viewed from the planet in four meridians of northern or southern phase $\Psi_{N\!,S}$.
  The sense of the principal PPO currents are shown by the green lines and arrows, closing along field lines out of the plane of the page.
  Purple solid and dashed lines show the location and tilt of the cross-tail current and its north-south component, respectively.
  The corresponding orientation of the PPO core field is shown by the blue arrows and icons.
  }
  \label{fig:tail}
\end{figure}

The significant discrepancy between the total PPO associated north-south currents measured here on the nightside and those reported from studies of high-latitude data may be explained by periodic motion of the nightside plasma sheet and the cross-tail current, flowing in the sense of planetary rotation, principally from dusk to dawn across the tail.
Modulations of the both the thickness and the vertical location of the plasma sheet have been shown to be well organised by the phases of the northern and southern PPO systems~\citep[e.g.][]{morooka09a, arridge11a, thomsen17a}.
The superposition of the PPO fields is such that the effective center of the plasma sheet, i.e.\ the location where the quasi-static radial field reverses sense, undergoes a maximal displacement towards the north at $\Psi_{N\!,S}=0^\circ$ and towards the south at $\Psi_{N\!,S}=180^\circ$.
At intermediate phases, the tilts maximize from north to south at 90$^\circ$ and from south to north at 270$^\circ$, giving a southward contribution to the current at 90$^\circ$ and a northward contribution at 270$^\circ$, just as do the interhemispheric currents depicted in Figure~\ref{fig:ppo}.
This periodic tilting of the plasma sheet is illustrated schematically in Figure~\ref{fig:tail}, as viewed from Saturn in four meridians of phase $\Psi_{N,S}$.
The PPO modulated interhemispheric currents are shown by the green lines and arrows, and close along planetary field lines out the plane of the page.
The plasma sheet current is shown by the solid purple line, flowing in the sense of planetary rotation, and where tilted its north-south component is indicated by the arrowed dashed lines, which can be seen to flow in the same direction as the interhemispheric currents.
To aid comparison with Figure~\ref{fig:ppo}, in Figure~\ref{fig:tail} we also depict the orientation of the core field above and below the current sheet equator.

The day-night differences evident in the results presented here then suggest that these effects are much more pronounced in the stretched out magnetodisk and tail field regions on the nightside (where they have been principally observed) than in the more compressed field region of the dayside magnetosphere.
Hence, the four panels shown in Figure~\ref{fig:tail} can be viewed as representing the configuration of fields and currents in the tail as viewed from the planet, with dawn to the left and dusk to the right.
As a PPO system rotates around the planet, an additional northward-directed current then appears near dusk as the $\Psi_{N\!,S}=270^\circ$ meridian passes through that sector (corresponding to $\Phi_{N\!,S}=0^\circ$ as shown in Figure~\ref{fig:ppo}), rotates with that meridian across midnight (corresponding to $\Phi_{N\!,S}=90^\circ$, with appropriate phase delays depending on radial distance), and then fades again as that meridian passes through dawn (corresponding to $\Phi_{N\!,S}=180^\circ$).
At the same time as the latter, the $\Psi_{N\!,S}=90^\circ$ meridian is passing through the dusk, where a growing reversed tilt of the current sheet leads to a growing region of enhanced southward current flowing through the equatorial plane, which similarly rotates via midnight (corresponding to $\Phi_{N\!,S}=270^\circ$), and fades near the dawn meridian (corresponding to $\Phi_{N\!,S}=0^\circ$).
Such behavior seems qualitatively in accord with that deduced here (and by~\citet{andrews10a}) and shown in Figure~\ref{fig:cur}.
More quantitatively, a reasonable estimate of the current density of the cross-tail current at radial distances of 15-30~\RS\ on the nightside is $j_{T}\approx25\;\mathrm{pA\;m^{-2}}$, obtained from inspection of the magnitude and length scale of the reversal of the field across the current layer.
A typical tilt angle of the plasma sheet center of $10^\circ$, in line with values found based on studies of field and particle data~\citep[e.g.][]{morooka09a, arridge11a, szego13a} thus suggests a PPO-related contribution of the ring current/cross-tail current to the measurements here of amplitude $j_{Tppo}\approx j_T\sin10^\circ\approx5\;\mathrm{pA\;m^{-2}}$.
Such a current, in phase with the principal PPO-related currents, therefore corresponds to $\sim$50\% of the measured peak north-south current density found on the nightside as shown in Figure~\ref{fig:cur}, or $\sim$2~MA.
This represents a modest fraction of the $\sim$9~MA expected current flowing cross-tail in the interval from 15 to 30~\RS, based on an empirical model of the tail field~\citep{jackman11a}.
Conversely, on the basis of results presented by~\citet{bradley18a}, any PPO-related modulation of the ring current itself is expected to be of the order of $\sim$0.5~MA in amplitude, and therefore only a small fraction of the quasi-static current of $\sim$10~MA.
The seasonal `warping' of Saturn's plasma sheet, as described e.g.\ by~\citet{arridge08a} and~\citet{carbary16b}, deflecting its central location to the northern hemisphere during southern summer and vice-versa may modify some elements of this discussion.

The combined modulation of the plasma sheet by both PPO systems leads, under certain conditions, to significantly non-sinusoidal variations in its location and thickness~\citep{szego13a, thomsen17a, cowley17a}, with this effect having been modelled both by refinement of the theoretical description of the core PPO field perturbations~\citep{cowley17b} and numerically~\citep{jia12c}.
The approach taken in this paper, determining common spatial variations of  phase and amplitude of the field oscillations relative to both `core' region fields, is such that effects associated with dual-modulated motion of the central plasma sheet are consistently treated.

The improved coverage of the equatorial plane obtained in this study allows meaningful comparisons between the properties of the PPO oscillations at different LTs.
For example, a significant day-night asymmetry is observed, with typically weaker amplitude oscillations observed on the dayside in all components, and consequently weaker current densities as compared to equivalent radial distances on the nightside.
This asymmetry is likely the result of the larger influence of the periodic motion of the tail current sheet on the nightside, discussed above.
Furthermore, this day-night asymmetry in current density has the effect of appearing to distort the structure of the PPO currents, such that the maximum upward and downward currents at a given phase are no longer directly opposite in LT.
Aperiodic magnetic field perturbations induced by motion of the magnetopause in response to solar wind variations likely also introduce appreciable noise to the analysis presented here, leading to a net underestimate of the PPO field amplitude at large distances on the dayside.
Similarly, a pronounced dawn-dusk asymmetry is also present in the results presented here, most notably in that the amplitude of the radial component of the oscillation is maximal in a narrow band of LT centered on $\sim$21~h, extending from the edge of the core region at $\sim$12~\RS\ to the outer boundary of this study at 30~\RS.
How this asymmetry in the relative amplitudes of the PPO magnetic fields may relate to similar asymmetries present in the quasi-static magnetospheric magnetic fields in the same region, or dynamical processes in the magnetotail and the interaction with the solar wind, remains to be fully explored.
Several studies have elucidated the role played by PPO fields in dynamic processes in Saturn's magnetotail, for example through modifying the properties of the tail current sheet and favouring the growth of plasma instabilities, magnetic reconnection and the release of plasmoids down-tail at certain PPO phases~\citep[e.g.,][]{arridge11a, jackman09b, jackman16a}.
Recently, through a large statistical survey of reconnection events in Saturn's tail, \citet{bradley18b} have shown that a high degree of organisation is indeed evident when considering specifically the PPO phase at the assumed reconnection site itself, rather than at the location of the observed plasmoid signature.

The effect of significant contributions to the north-south currents determined in this study by periodic plasma sheet motions must also be considered when comparing these results to auroral observations.
In a recent analysis of Cassini UVIS observations, \citet{bader18a} found that UV emissions from the northern hemisphere peaked at $\Psi_N=114^\circ$, thus close to the expected phase of maximum upward current in the northern ionosphere ($\Psi_N=90^\circ$) on the basis of the simple theoretical model.
\citet{bader18a} show that the peak UV brightness is found in the pre-noon sector, consistent with related observations of SKR emissions~\citep[e.g., ][]{lamy09a}, while no corresponding enhancement is present in the pre-noon sector in the equatorial currents depicted in Figure~\ref{fig:cur}.
As we suggest, a significant fraction of the north-south current shown in Figures~\ref{fig:cur} and~\ref{fig:cur_ph} in the post-dusk to pre-dawn sectors is produced by periodic tilting of the tail current sheet, and therefore does not close along the field lines via the auroral ionosphere.
Were it possible to accurately remove this component of the current, the remainder would likely be more constant in amplitude with LT, and thus closer to that expected from the simple theoretical model.
Such a current system would also be more consistent with the results presented by~\cite{hunt16a}, where PPO currents observed on high-latitude field lines were found to have only a small variation in amplitude with LT.
A quasi-sinusoidal variation of the PPO currents with LT at a given oscillation phase and radial distance is moreover required to produce the observed quasi-uniform magnetic field perturbation in the core region.

\citet{andrews10a}, in their earlier analysis of equatorial PPO field structures, noted an apparent exclusion effect in the innermost regions sampled.
Currents flowing field-aligned at and interior to this region were inferred to largely screen out the PPO fields from the smallest radial distances, suggested to be a requirement of the presence of a highly conducting E-ring plasma.
However, during the latter part of the Cassini mission, the periapsis of the spacecraft during the so-called `F-ring' and `proximal' intervals reached smaller distances than previously attained (excepting briefly during Saturn orbit insertion).
During the F-ring orbits,~\citet{hunt18a} observed field-aligned currents in high-latitude magnetic field data that mapped to distances of $\sim$2 to~7~\RS\ in the equatorial plane, and were modulated with both northern and southern PPO phase, well inside the E-ring plasma torus.
It remains to be seen whether the PPO fields, are, or can be shown to be present on field lines both threading and interior to Saturn's rings.

Finally, following~\citet{provan16a}, we speculate on the nature of the PPO fields and currents during the apparently unique interval from mid 2013 to mid 2014 within which the rotation periods of the northern and southern systems coalesced.
Constraints imposed by Cassini's orbit during this interval precluded the possibility of determining the ratio of the two system amplitudes $k$, although~\citet{provan16a} tentatively suggested a value of $k\approx1$ on the basis of measurements made immediately preceding the coalescence of the two periods.
Crucial then is the relative phase of the two systems $\Delta\Phi = \Phi_N - \Phi_S$.
Based on the results provided by~\citet{provan16a}, $\Delta\Phi$ was $\sim$180$^\circ$ during this interval, and therefore, with approximately equal amplitudes, the transverse components of the two fields remain opposite and cancel one-another on the equator.
Consequently, the net field-aligned currents associated with the PPO fields during this interval are expected to be vanishingly small at the equator, and the transfer of angular momentum between the hemispheres marginal.

\section{Summary}\label{sec:summary}
In this study we have revisited the determination of the spatial structure of planetary-period oscillations in Saturn's equatorial magnetic field, thereby quantifying departures from the simple theoretical model proposed initially by~\citet{southwood07a} and~\citet{andrews08a}.
In doing so, we extend the previous related study by~\citet{andrews10a} in two important respects.

Firstly, we use all suitable magnetic field data taken throughout the Cassini mission, spanning four extended intervals of near-equatorial orbits.
The total volume of data analysed in this study is thereby increased by $\sim$350\% over that previously processed by~\citet{andrews10a}, and the consequent improvement in spatial coverage is such that the equatorial plane is now essentially fully sampled over radial distances from $\sim$3 to 30~\RS.
Only a few small gaps in coverage remain, in which a suitable volume of data were not obtained during the Cassini orbital tour.
In particular, significantly improved coverage is obtained in the afternoon and pre-dusk sectors compared to the earlier analysis by~\citet{andrews10a}, allowing meaningful investigations of dawn-dusk asymmetries of the relative amplitude and phase of the PPO fields and currents.
Furthermore, the sampling of the equatorial magnetosphere is such that we can be confidant that no major systematic biases remain in the analysis, such as could lead, e.g.\, to favouring observations in one particular sector of LT within only specific, restricted seasonal intervals.
Nevertheless, the nature of Cassini's orbital tour, combined with the seasonal warping of the Saturnian plasma sheet is such that more data are obtained below the magnetospheric equator than above, although the magnitude of this bias is reduced compared to that present in the original study by~\citet{andrews10a}.

Secondly, the analysis presented here takes full account of the presence of two superposed systems of field oscillations, each associated with a specific hemisphere.
The original study by~\citet{andrews10a} only considered the effects of the (then dominant) southern hemisphere system.
Here, independently varying northern and southern system rotation periods determined primarily from core-region (dipole $L < 12$~\RS) magnetic field data are used to define the instantaneous rotation phase of each system.
In addition, we consider the piece-wise determined secular changes in the amplitudes of the individual components of the core field for both the northern and southern systems.
Both the rotation periods and the amplitudes of the two systems are obtained in a series of related papers spanning all of Cassini's equatorial orbit intervals, and synthesised for the purposes of this analysis as shown in Figure~\ref{fig:ts}~\citep{andrews12a, provan13a, provan16a, provan18a}.
The time-varying amplitudes and rotation periods of both systems are used to define a simple superposition of both quasi-uniform PPO fields in the `core' region, against which common spatial variations in relative amplitudes and phases are determined for all three components of the magnetic field.
From this determination of common variations, phase variations associated with the modulations in the northern and southern hemispheres can be separately constructed via the addition of a constant phase offset appropriate for the hemisphere and component in question.
Fully determining the spatial variations in relative amplitude and phase of the transverse equatorial fields affords the possibility of determining the component of the electrical current flowing north-south through the equatorial plane.

We now briefly summarise the central results of this study.
\begin{itemize}

  \item An initial analysis was first performed utilising a simplified method, similar to that used by~\citet{andrews10a}, treating only the PPO fields associated with a single hemisphere in isolation.
  Since the phases of the two PPO systems are overall uncorrelated, given sufficient data within each spatial bin, both in terms of volume and distribution in time, perturbations induced by the simultaneously present field of the opposing hemisphere will mainly contribute ``noise'' to the analysis, and will only marginally contribute to the amplitude and phase determined in each case.
  Results of this analysis, presented in section~\ref{sub:indep}, are clearly suggestive of a spatial structure of the relative amplitudes and phases that is common to both the northern and southern systems, apart from the reversal in sign of the colatitudinal field component, hence justifying the `combined-fitting' approach taken in the remainder of the results presented in this paper.
  No systematic evidence was found for differences between the two hemispheres on large scales.
  That there exists a common spatial variation of the field oscillations in both systems is strongly suggestive of a common phenomenon driving both systems, albeit with amplitudes and phases that may differ at a given time.

  \item The large-scale structure of the equatorial field oscillations, common to both the northern and southern PPO systems as subsequently determined in this paper, is consistent with the results reported first by~\citet{andrews10a} (Figures~\ref{fig:str}-\ref{fig:cur}).
  Specifically, the quasi-uniform core field is bounded by regions of net circulation, indicative of north-south currents flowing at $\sim$15~\RS, which are generally most intense in the post-dusk to post-dawn sectors.
  At larger distances, the amplitude of the PPO fields generally decreases, most rapidly on the dayside, while the field itself takes on the form of a rotating transverse dipole, to first-order.
  The observed structure of the PPO field therefore remains in good agreement with the description provided by the simple axisymmetric models proposed by~\citet{southwood07a} and \citet{andrews08a}.

  \item Nevertheless, substantial variations in both amplitude and phase are present in all magnetic field components with both radial distance and LT (Figure~\ref{fig:str}), leading to appreciable departures from the simple axisymmetric models.
  These variations are typically smallest within the core region, but nevertheless are measurable here, amounting to steady increases in relative phase with radial distance, and somewhat smaller amplitudes in the innermost spatial bins sampled.
  More marked LT variations in amplitude are seen in all components outside the core region, where field amplitudes are typically larger on the nightside than on the dayside, whilst falling with radial distance.
  However, this general trend is not in evidence for the radial component, in which the amplitudes remain close to their maximum both in a band of LT centered on radial distances of $\sim$15~\RS\ and spanning $\sim$18 - 09~h~LT (via midnight), and in a connected region extending outward to at least $\sim$30~\RS\ at LTs centered on 21~h.
  Furthermore, the relative phase of the azimuthal component is seen to rapidly increase by $\sim$180$^\circ$ across radial distances centered on $\sim$15~\RS\ within an extended range of LT spanning the pre-midnight to pre-noon sectors.
  Meanwhile, for the radial component, relative phases in the afternoon sector vary only weakly with increasing radial distance, indicative of much faster outward propagation of the perturbations in this region.
  A clear dawn-dusk asymmetry is thus present, and is largest in all respects in the transverse components of the PPO fields.
  This asymmetry extends to the currents flowing through the equatorial plane, which are confined to a much narrower range of radial distance in the dawn sector compared to the dusk sector.
  Such a dawn-dusk asymmetry in the north-south currents measured on the equatorial plane may lead to a related dawn-dusk asymmetry in the  modulation of the auroras by the PPO systems.
  The colatitudinal component shows markedly smaller variations in amplitude and phase with LT than both the radial and azimuthal components.

  \item Radial phase velocities, computed from the gradient in phase with radial distance, are typically in the range $\sim$100-300~$\mathrm{km\;s^{-1}}$ across all field components.
  These values are computed at all radial distances for the radial $r$ and colatitudinal $\theta$ components, and inside of 15~\RS\ for the azimuthal $\varphi$ component (thereby avoiding the `structural' shear in $B_\varphi$ associated with the principal field aligned currents).
  We note that a region of significantly larger phase velocities, $\sim$800$\mathrm{km\;s^{-1}}$, is present in the radial component data in the afternoon sector, 12 - 18~h LT.

  \item Azimuthal phase velocities, computed from estimates of the amplitude of $m=1$ variations in relative phases $\xi_i$ with LT, show only modest departures from co-rotation with the core PPO rotation rate in the colatitudinal component.
  However, for the transverse $r-\varphi$ field components, the velocities are such that phase fronts significantly lead the core rotation throughout most of the dayside, the angular velocity typically peaking slightly pre-noon, whilst being retarded at other LTs.

  \item A coherent picture emerges in which two principal regions of current, one of each sign, are present at all rotation phases (Figure~\ref{fig:cur}).
  Each current region forms a near-continuous spiral, from the innermost radial distances sampled, and outwards to distances beyond the `core' region at progressively earlier LTs.
  The current density of a given region of current typically peaks as it crosses radial distances of $\sim$15~\RS\ in the pre-midnight to post-dawn sectors.
  These two spirals of north-south current are interlinked at all rotation phases, such that at a given LT, there is generally a reversal in sign of the current at a radial distance $\sim$12~\RS.

  This interpretation thus represents a simplification compared to the situation described by~\citet{andrews10a}, based on their early analysis of a sub-set of these data.
  Considering the averaged behaviour over the interval studied, the total current flowing through the equatorial plane associated with the southern PPO system varies with phase over the range $\sim$4.0 to $\sim$5.1~MA (Figure~\ref{fig:cur_ph}).
  Of this total, the so-called `principal' currents flowing outside of the core $r>$12~\RS\ carry $\sim$3.1 to $\sim$4.5~MA, whilst those inside 12~\RS\ contribute $\sim$0.5-1.1~MA.

  \item Comparing the inferred north-south currents on the equatorial plane with those estimated from separate analysis of high-latitude magnetic field data by~\citet{hunt14a, hunt15a} and~\citet{bradley18a}, we note a 25-50\% discrepancy in their magnitudes.
  This discrepancy is however readily explained by considering the contribution to the north-south current measured on the equatorial plane due to PPO-related vertical motions of the center of Saturn's nightside tail current sheet.
  This effect appears principally constrained to larger radial distances, being most evident beyond $\sim$15~\RS\ on the nightside.
  We show in Figure~\ref{fig:tail} that the north-south component of the total current flowing within the sheet is modulated in phase with the principal PPO currents, and thus increases the strength of the measured north-south current on the nightside by an amount sufficient to explain the observed discrepancy with high-latitude measurements.
  The additional contribution of the periodic tilting of the tail current sheet to the measured equatorial currents is also sufficient to at least partially obscure the underlying sinusoidal nature of the principal PPO currents associated with each hemisphere.
  No corresponding evidence is found for the presence of a similar effect at smaller radial distances and on the dayside associated with periodic displacement of the ring current.

  \item  Considering the fields of the northern and southern systems combined at specific intervals within the Cassini mission, representative of intervals of either strong southern or northern dominance, or near-equal amplitudes, estimates for the total current flowing through the equatorial plane have been made (Figure~\ref{fig:dual}).
  Extremal values are associated with intervals of near-equal amplitudes of the two systems, where PPO fields and currents almost completely reinforce or destroy one another depending on the instantaneous value of the beat phase $\Delta\Phi$.
  During these intervals, the total current reaches a maximum of $\sim$11~MA at $\Delta\Phi=0^\circ$ and a minimum of $\sim$0.1~MA at $\Delta\Phi=180^\circ$.

\end{itemize}

\section*{Acknowledgements}
Cassini magnetometer data used in this study is available from the NASA Planetary Data System: \url{https://pds.nasa.gov}.
DJA was supported by grant DNR:162/14 from the Swedish National Space Agency.
Work at Leicester and Imperial College was supported by STFC Consolidated Grants ST/N000749/1 and ST/N000692/1, respectively.
This study benefitted from discussions at the International Space Science Institute, within the international team on ``Rotational phenomena in Saturn's magnetosphere.''


\end{document}